\documentclass[12pt,a4paper]{article}
\usepackage[utf8]{inputenc}
\usepackage{amsmath}
\usepackage{amsfonts}
\usepackage{amssymb}
\usepackage{graphicx}
\usepackage{rotating}
\usepackage{pdflscape}
\usepackage{xspace}
\usepackage{xcolor}
\usepackage{cite}
\usepackage{mciteplus}
\usepackage{upgreek}
\usepackage{hyperref}
\usepackage[all]{hypcap}
\usepackage{ifthen}
\usepackage{authblk}

\usepackage{lineno}

\newboolean{uprightparticles}
\setboolean{uprightparticles}{false} 

\newboolean{articletitles}
\setboolean{articletitles}{true}

\def\MagUp {\mbox{\em Mag\kern -0.05em Up}\xspace}

\ifthenelse{\boolean{uprightparticles}}%
{

 \def\Ppi         {\ensuremath{\uppi}\xspace}

 \def\PDelta      {\ensuremath{\Delta}\xspace}                 
 \def\PXi      {\ensuremath{\Xi}\xspace}                 
 \def\PLambda      {\ensuremath{\Lambda}\xspace}                 
 \def\PSigma      {\ensuremath{\Sigma}\xspace}                 
 \def\POmega      {\ensuremath{\Omega}\xspace}                 
 \def\PUpsilon      {\ensuremath{\Upsilon}\xspace}                 
 
 \def\PB      {\ensuremath{\mathrm{B}}\xspace}                 
                  
 \def\PD      {\ensuremath{\mathrm{D}}\xspace}

 \def\PK      {\ensuremath{\mathrm{K}}\xspace}

 \def\Pb      {\ensuremath{\mathrm{b}}\xspace}                 
 \def\Pc      {\ensuremath{\mathrm{c}}\xspace}

 \def\Pi      {\ensuremath{\mathrm{i}}\xspace}

 \def\Pu      {\ensuremath{\mathrm{u}}\xspace}

}
{

 \def\Ppi         {\ensuremath{\pi}\xspace}

 \mathchardef\PDelta="7101
 \mathchardef\PXi="7104
 \mathchardef\PLambda="7103
 \mathchardef\PSigma="7106
 \mathchardef\POmega="710A
 \mathchardef\PUpsilon="7107
                  
 \def\PB      {\ensuremath{B}\xspace}                 
                  
 \def\PD      {\ensuremath{D}\xspace}

 \def\PK      {\ensuremath{K}\xspace}

 \def\Pb      {\ensuremath{b}\xspace}                 
 \def\Pc      {\ensuremath{c}\xspace}

 \def\Pi      {\ensuremath{i}\xspace}

 \def\Pu      {\ensuremath{u}\xspace}

}

\makeatletter
\ifcase \@ptsize \relax
  \newcommand{\miniscule}{\@setfontsize\miniscule{4}{5}}
\or
  \newcommand{\miniscule}{\@setfontsize\miniscule{5}{6}}
\or
  \newcommand{\miniscule}{\@setfontsize\miniscule{5}{6}}
\fi
\makeatother

\DeclareRobustCommand{\optbar}[1]{\shortstack{{\miniscule (\rule[.5ex]{1.25em}{.18mm})}
  \\ [-.7ex] $#1$}}

\def\uquark    {{\ensuremath{\Pu}}\xspace}

\def\cquark    {{\ensuremath{\Pc}}\xspace}

\def\bquark    {{\ensuremath{\Pb}}\xspace}

\def\pion   {{\ensuremath{\Ppi}}\xspace}

\def\pip    {{\ensuremath{\pion^+}}\xspace}
\def\pim    {{\ensuremath{\pion^-}}\xspace}

\def\kaon    {{\ensuremath{\PK}}\xspace}
  \def\Kbar    {{\kern 0.2em\overline{\kern -0.2em \PK}{}}\xspace}

\def\KorKbar    {\kern 0.18em\optbar{\kern -0.18em K}{}\xspace}

\def\Kp      {{\ensuremath{\kaon^+}}\xspace}
\def\Km      {{\ensuremath{\kaon^-}}\xspace}

\def\KS      {{\ensuremath{\kaon^0_{\rm\scriptscriptstyle S}}}\xspace}
\def\KL      {{\ensuremath{\kaon^0_{\rm\scriptscriptstyle L}}}\xspace}

  \def\Dbar    {{\kern 0.2em\overline{\kern -0.2em \PD}{}}\xspace}
\def\D       {{\ensuremath{\PD}}\xspace}

\def\DorDbar    {\kern 0.18em\optbar{\kern -0.18em D}{}\xspace}
\def\DtwoorDtwobar {\kern -0.25em\optbar{\kern 0.25em D_2^*}{}\xspace}
\def\Dz      {{\ensuremath{\D^0}}\xspace}
\def\Dzb     {{\ensuremath{\Dbar{}^0}}\xspace}

\def\B       {{\ensuremath{\PB}}\xspace}
\def\Bbar    {{\ensuremath{\kern 0.18em\overline{\kern -0.18em \PB}{}}}\xspace}

\def\BorBbar    {\kern 0.18em\optbar{\kern -0.18em B}{}\xspace}
\def\Bz      {{\ensuremath{\B^0}}\xspace}

\def\BzorBzbar  {\kern 0.18em\optbar{\kern -0.18em B}{}^0\xspace}
\def\Bu      {{\ensuremath{\B^+}}\xspace}
\def\Bub     {{\ensuremath{\B^-}}\xspace}
\def\Bp      {{\ensuremath{\Bu}}\xspace}
\def\Bm      {{\ensuremath{\Bub}}\xspace}

  \def\Y#1S{\ensuremath{\PUpsilon{(#1S)}}\xspace}

\def\Lbar        {{\ensuremath{\kern 0.1em\overline{\kern -0.1em\PLambda}}}\xspace}
\def\LorLbar    {\kern 0.18em\optbar{\kern -0.18em \PLambda}{}\xspace}

\def\to                 {\ensuremath{\rightarrow}\xspace}

\def\CP                {{\ensuremath{C\!P}}\xspace}

\def\AT#1     {\ensuremath{A_{\mathrm{T}}^{#1}}\xspace}           

\def\C#1      {\ensuremath{\mathcal{C}_{#1}}\xspace}                       
\def\Cp#1     {\ensuremath{\mathcal{C}_{#1}^{'}}\xspace}                    
\def\Ceff#1   {\ensuremath{\mathcal{C}_{#1}^{\mathrm{(eff)}}}\xspace}        
\def\Cpeff#1  {\ensuremath{\mathcal{C}_{#1}^{'\mathrm{(eff)}}}\xspace}       
\def\Ope#1    {\ensuremath{\mathcal{O}_{#1}}\xspace}                       
\def\Opep#1   {\ensuremath{\mathcal{O}_{#1}^{'}}\xspace}                    

\newcommand{\tev}{\ifthenelse{\boolean{inbibliography}}{\ensuremath{~T\kern -0.05em eV}\xspace}{\ensuremath{\mathrm{\,Te\kern -0.1em V}}}\xspace}
\newcommand{\gev}{\ensuremath{\mathrm{\,Ge\kern -0.1em V}}\xspace}
\newcommand{\mev}{\ensuremath{\mathrm{\,Me\kern -0.1em V}}\xspace}
\newcommand{\kev}{\ensuremath{\mathrm{\,ke\kern -0.1em V}}\xspace}
\newcommand{\ev}{\ensuremath{\mathrm{\,e\kern -0.1em V}}\xspace}
\newcommand{\gevc}{\ensuremath{{\mathrm{\,Ge\kern -0.1em V\!/}c}}\xspace}
\newcommand{\mevc}{\ensuremath{{\mathrm{\,Me\kern -0.1em V\!/}c}}\xspace}
\newcommand{\gevcc}{\ensuremath{{\mathrm{\,Ge\kern -0.1em V\!/}c^2}}\xspace}
\newcommand{\gevgevcccc}{\ensuremath{{\mathrm{\,Ge\kern -0.1em V^2\!/}c^4}}\xspace}
\newcommand{\mevcc}{\ensuremath{{\mathrm{\,Me\kern -0.1em V\!/}c^2}}\xspace}

\def\gsim{{~\raise.15em\hbox{$>$}\kern-.85em
          \lower.35em\hbox{$\sim$}~}\xspace}
\def\lsim{{~\raise.15em\hbox{$<$}\kern-.85em
          \lower.35em\hbox{$\sim$}~}\xspace}

\def\tell1  {TELL1\xspace}
\def\ukl1   {UKL1\xspace}

\newcommand{\ie}{\mbox{\itshape i.e.}\xspace}

\newcommand{\dvar}{\ensuremath{\mathbf{z}}\xspace}
\newcommand{\DP}{\ensuremath{\mathcal{D}}\xspace}
\newcommand{\dkpp}{\ensuremath{D\to \KS\pip\pim}\xspace}
\newcommand{\dnkpp}{\ensuremath{\Dz\to \KS\pip\pim}\xspace}
\newcommand{\dbkpp}{\ensuremath{\Dzb\to \KS\pip\pim}\xspace}
\newcommand{\kpp}{\ensuremath{\KS\pip\pim}\xspace}
\newcommand{\dkkk}{\ensuremath{\Dz\to \KS\Kp\Km}\xspace}
\newcommand{\bdk}{\ensuremath{B\to DK}\xspace}
\newcommand{\bpmdk}{\ensuremath{B^{\pm}\to DK^{\pm}}\xspace}
\newcommand{\bpdk}{\ensuremath{\Bp\to D\Kp}\xspace}
\newcommand{\bmdk}{\ensuremath{\Bm\to D\Km}\xspace}
\newcommand{\bdkpi}{\ensuremath{B\to DK\pi}\xspace}
\newcommand{\ddb}{\ensuremath{\Dz\Dzb}\xspace}

\title{Unbinned model-independent measurements with 
       coherent admixtures of 
       multibody neutral $D$ meson decays}
\author{Anton Poluektov}
\affil{\small Department of Physics, University of Warwick, Coventry, United Kingdom}
\date{\small 22 December 2017}

\begin{document}


\maketitle

\begin{abstract}
  \noindent
  Various studies of Standard Model parameters involve measuring the properties 
  of a coherent admixture of \Dz and \Dzb states. 
  A typical example is the determination of the Unitarity
  Triangle angle $\gamma$ in the decays \bdk, \dkpp. A model-independent approach to 
  perform this measurement is proposed that has superior statistical sensitivity than the 
  well-established method involving binning of the \dkpp decay phase space. The technique 
  employs Fourier analysis of the complex phase difference between \Dz and \Dzb 
  decay amplitudes and can be easily generalised to other similar measurements, 
  such as studies of charm mixing or determination of the angle $\beta$ from $\Bz\to D h^0$ decays. 
\end{abstract}

\section{Introduction}

\label{sec:introduction}

Precise measurements of \CP violation in decays of beauty hadrons is one of the key methods  to search for effects of physics beyond the Standard Model. The phenomenon of \CP violation is described in the Standard Model (SM) by the Cabibbo-Kobayashi-Maskawa (CKM) mechanism~\cite{Cabibbo:1963yz,Kobayashi:1973fv}, where \CP violation enters as a complex phase 
in the unitary $3\times 3$ matrix (CKM matrix) describing transitions between quarks 
of the three generations due to charged-current weak interactions. A common representation of the CKM matrix is the Unitarity Triangle (UT), the sides and angles of which are experimentally observable parameters. The fundamental \CP-violating phase, the angle $\gamma$ of the UT (also known in the literature as $\phi_3$), can be obtained with extremely low theoretical uncertainty~\cite{Brod:2013sga} from tree-dominated $\bquark$ hadron decays and thus serves as a ``standard candle'' for searches of effects beyond the Standard Model in other heavy flavour processes. 

Various techniques have been proposed to measure $\gamma$ experimentally in the decays of $B$ mesons into final states with neutral $D$ mesons~\cite{Gronau:1990ra,Gronau:1991dp,Atwood:1996ci,Atwood:2000ck}. The \CP violation in these decays is generated by interference of $\bquark\to\cquark$ and $\bquark\to\uquark$ quark level transitions once the neutral $D$ meson is reconstructed in a final state accessible to both \Dz and \Dzb decays. The neutral $D$ meson in this case forms a coherent admixture of $\Dz$ and $\Dzb$ states which is denoted here as $D$. 
One of the most sensitive techniques involves analysis of the Dalitz plot density of multibody $D$ decays such as \dkpp~\cite{Giri:2003ty,Bondar}. 

Two different techniques have been developed and implemented experimentally 
to extract $\gamma$ from \bdk decays using multibody 
$D$ meson final states. 
One is model-dependent, with the complex amplitude of the $D$ decay obtained by fitting the flavour-specific $\Dz$ decay density to a model~\cite{Poluektov:2004mf,Aubert:2005iz,Poluektov:2006ia,Aubert:2008bd,Poluektov:2010wz,Aaij:2014iba,AbellanBeteta:2016zlt}. 
This technique offers optimal statistical precision since the fit can be performed 
in an unbinned fashion, however, it suffers from uncertainty, which is difficult to quantify, 
due to modelling of the $\Dz$ amplitude. Another method is a binned model-independent approach, 
where information about the behaviour of the strong phase across the phase space 
of the \Dz decay is obtained from samples of quantum-correlated \ddb decays produced
near kinematic threshold~\cite{Giri:2003ty,Bondar:2005ki,Bondar:2008hh,Aaij:2014uva,Aaij:2012hu}. 

In the model-independent technique, one needs to determine the relation between the decay densities of 
quantum-correlated \ddb decays and $D$ decays from \bdk. This necessarily 
requires estimation of the decay density from scattered data, which is achieved by binning both 
decay densities. Each bin is assigned a number of parameters that characterise the averaged
behaviour of the amplitude (its magnitude and phase) over the bin; 
these parameters are obtained by solving a 
system of equations that also includes the value of $\gamma$. In general, the binned approach 
reduces statistical sensitivity compared to the unbinned model-dependent technique, 
but the procedure is developed in such a way that it produces an unbiased 
measurement even in the case of a very rough binning. 

In this paper, a method to extract $\gamma$ is proposed which does not involve binning and aims to combine the advantages of the model-dependent and model-independent approaches. 
Like the binned approach with optimal binning, it uses a construction inspired by a
$\Dz$ amplitude model, but provides an unbiased measurement even if the wrong model is used. 
It is shown to offer better statistical sensitivity than the binned approach. 
The method employs Fourier analysis of the distribution of the complex phase difference 
between the \Dz and \Dzb amplitudes. 
The method is illustrated using the ``golden'' channel \bdk with subsequent \dkpp decay, 
but can easily be generalised to other cases of $\gamma$ 
determination where the binned model-independent technique is applicable: analyses using other 
three- or four-body $\Dz$ decays~\cite{Artuso:2012df, Insler:2012pm, Aaij:2014dia, 
Aaij:2015jna, Harnew:2017tlp, K:2017qxf}, multibody $B$ 
decays~\cite{Negishi:2015vqa, Aaij:2016nao} or analyses 
using correlated Dalitz plots of multibody $B$- and $D$-meson decays~\cite{Gershon:2009qr, Craik}. 

Apart from measurements of $\gamma$, similar model-independent techniques, which employ interference between 
$\Dz$ and $\Dzb$ amplitudes, have been developed for other kinds of measurements: studies of \CP violation and mixing parameters of $\Dz$ mesons~\cite{Bondar:2010qs,Thomas:2012qf,Aaij:2015xoa}, measurements of the UT angle $\beta$ in $\Bz\to Dh^0$ (where $h^0$ is a neutral light meson) and $\Bz\to D\pip\pim$ decays~\cite{Bondar:2005gk,Vorobyev:2016npn}. In all these cases, the technique proposed can be applied instead of the binned methods. 

\section{Model-independent formalism with weight functions}

\label{sec:general}

In this section, the formalism for $\gamma$ measurement is recalled to introduce the notation, and 
the established model-independent technique is reformulated in slightly different terms. 
This allows a demonstration that the binned approach is not the only possible method to perform such a measurement. 

Measurements of $\gamma$ based on \bdk processes use the fact that the decay involves the interference of tree-dominated $\bquark\to\cquark$ and $\bquark\to\uquark$ diagrams, which produce neutral $D$ mesons with opposite flavours. In the case of \bpdk decays followed by \dkpp, the amplitude as a function of two variables of the $D$ decay Dalitz plot, the 
squared invariant masses $m^2_+ \equiv  m^2_{\KS\pip}$ and $m^2_-\equiv m^2_{\KS\pim}$, 
is expressed as 
\begin{equation}
  \overline{A}_{B}(m^2_+, m^2_-) = \overline{A}_D(m^2_+, m^2_-) + r_Be^{i(\delta_B+\gamma)} A_D(m^2_+, m^2_-), 
\end{equation}
where the first term is due to $\bar{\bquark}\to\bar{\cquark}$ and the second due to $\bar{\bquark}\to\bar{\uquark}$ transition. Here $A_D(m^2_+, m^2_-)$ is the amplitude of the \dnkpp decay, 
$\overline{A}_D(m^2_+, m^2_-)$ is that for the \dbkpp decay, $r_B$ is the relative magnitude of the two contributions and $\delta_B$ is the \CP-conserving strong phase between them. The amplitude $A_B$ for the \CP-conjugated decay \bmdk can be obtained by replacing $\gamma\to -\gamma$ and swapping the $D$ decay amplitudes: $\overline{A}_D\leftrightarrow A_D$. A simultaneous analysis of the two amplitudes $A_B$ and $\overline{A}_B$
provides information on the unknown parameters $\gamma$, $r_B$, and $\delta_B$. 

Experimentally, one deals with probability densities rather than amplitudes. The decay density for \bpdk decays as a function of $\dvar\equiv (m^2_+, m^2_-)$ is
\begin{equation}
  \bar{p}_{\B}(\dvar) \propto |\overline{A}_D(\dvar) + r_B e^{i\delta_B + i\gamma} A_D(\dvar)|^2 = 
                   |\overline{A}_D(\dvar) + (x_+ + iy_+) A_D(\dvar)|^2,
\end{equation}
where the Cartesian \CP-violating observables are introduced: $x_+ = r_B\cos(\delta_B+\gamma)$ and 
$y_+=r_B\sin(\delta_B+\gamma)$. The decay density $p_B(\dvar)$ for \bmdk decay involves the corresponding parameters $x_- = r_B\cos(\delta_B-\gamma)$ and $y_- = r_B\sin(\delta_B-\gamma)$: 
\begin{equation}
  p_{\B}(\dvar) \propto |A_D(\dvar) + r_B e^{i\delta_B - i\gamma} \overline{A}_{D}(\dvar)|^2 = 
                   |A_D(\dvar) + (x_- + iy_-) \overline{A}_D(\dvar)|^2. 
\end{equation}
The expressions for the decay densities can be rewritten as
\begin{equation}
  \begin{split}
  \bar{p}_{\B}(\dvar) & = \bar{h}_{\B}\left\{\bar{p}_D(\dvar) + r_B^2 p_D(\dvar) + 2[x_+C(\dvar) - y_+S(\dvar)]\right\}, \\
  p_{\B}(\dvar) & = h_{\B}\left\{p_D(\dvar) + r_B^2 \bar{p}_D(\dvar) + 2[x_-C(\dvar) + y_-S(\dvar)]\right\}, 
  \end{split}
  \label{eq:pb_dp}
\end{equation}
where $h_{\B}$ and $\bar{h}_{\B}$ are the normalisation factors, and 
$p_D(\dvar)$ and $\bar{p}_D(\dvar)$ are the Dalitz plot densities of 
flavour-tagged \dnkpp and \dbkpp decays:  
\begin{equation}
  p_D(\dvar) = |A_D(\dvar)|^2 = p_D(m^2_+, m^2_-), 
  \label{eq:pd_dp}
\end{equation}
\begin{equation}
  \bar{p}_D(\dvar) = |\overline{A}_D(\dvar)|^2 = p_D(m^2_-, m^2_+),
  \label{eq:pdbar_dp}
\end{equation}
{\it i.e.} the Dalitz plot distributions for $\Dz$ and $\Dzb$ decays are symmetric under the exchange $m^2_+\leftrightarrow m^2_-$ assuming \CP conservation in $\Dz$ decays.\footnote{The same assumption is made throughout this paper.}. 
The functions $C(\dvar)$ and $S(\dvar)$ 
contain information about the motion of the complex strong phase over the 
Dalitz plot which cannot be obtained from flavour-specific $D$ meson decays: 
\begin{equation}
  C(\dvar) = \mbox{Re}\left[A_D^*(\dvar)\overline{A}_D(\dvar)\right], \;\;\;
  S(\dvar) = \mbox{Im}\left[A_D^*(\dvar)\overline{A}_D(\dvar)\right]. 
\end{equation}
One needs to know them to obtain the values of \CP violating parameters $x_{\pm}$ and $y_{\pm}$ from $\bar{p}_{\B}(\dvar)$
and $p_{\B}(\dvar)$. 

In the model-dependent approach to measure $\gamma$, the strong phase motion is fixed by an amplitude model. The model-independent technique employs pairs of neutral $D$ mesons produced at the kinematic threshold in the $e^+e^-\to \ddb$ process to obtain this information. In this case, the two $D$ mesons are produced in a $P$-wave such that their wave function is antisymmetric. As a result, if both $D$ mesons are reconstructed in the \kpp final state, the densities of two Dalitz plots will be correlated:  
\begin{equation}
  \begin{split}
  p_{DD}(\dvar_1, \dvar_2) & = \frac{1}{2}|A_D(\dvar_1)\overline{A}_D(\dvar_2) - A_D(\dvar_2)\overline{A}_D(\dvar_1)|^2 = \\
     & h_{DD}\left\{p_D(\dvar_1)\bar{p}_D(\dvar_2) + p_D(\dvar_2)\bar{p}_D(\dvar_1) - 
       2 \left[C(\dvar_1) C(\dvar_2) + S(\dvar_1)S(\dvar_2) \right]\right\}. 
  \end{split}
  \label{eq:pdd_dp}
\end{equation}
Here the indices ``1'' and ``2'' correspond to the two decaying $D$ mesons and $h_{DD}$ is a normalisation factor. 
The necessary information about $C(\dvar)$ and $S(\dvar)$ is present in 
expression (\ref{eq:pdd_dp}), but it is not straightforward to obtain the explicit
expressions for the functions $C(\dvar)$ and $S(\dvar)$ from the observable distributions 
$p_D(\dvar)$, $\bar{p}_{D}(\dvar)$ and $p_{DD}(\dvar_1, \dvar_2)$. 

Equation~(\ref{eq:pdd_dp}) contains an ambiguity: it is invariant under rotation of the pair 
$C(\dvar)$, $S(\dvar)$ by an arbitrary phase $\Delta$: 
\begin{equation}
  \left(\begin{array}{c}C(\dvar) \\ S(\dvar)\end{array}\right) \rightarrow
  \left(\begin{array}{rr}\cos\Delta & \sin\Delta \\ -\sin\Delta & \cos\Delta\end{array}\right)
  \left(\begin{array}{c}C(\dvar) \\ S(\dvar)\end{array}\right). 
  \label{eq:ambig}
\end{equation}
This does not constitute a significant problem 
since it effectively results in the redefinition of the strong phase $\delta_B$, leaving the 
\CP-violating phase $\gamma$ unaffected. The other abiguity is the change of sign of $C(\dvar)$
or $S(\dvar)$, which results in the change of sign for $\gamma$. 
Other decays of $D$ mesons from correlated $\Dz\Dzb$ pairs 
can offer additional information to resolve these ambiguities. For instance, 
decays where one of the $D$ mesons is reconstructed in a \CP-eigenstate and the other is 
reconstructed as $\kpp$ constrain $C(\dvar)$, and resolve the ambiguity (\ref{eq:ambig}), as well as 
fix the sign for $C(\dvar)$. 
The remaining ambiguity, the sign of $S(\dvar)$, can be resolved by a weak model assumption using 
isobar parametrisation of the $D$ decay amplitude~\cite{Bondar:2008hh}. 
In practice, several $D$ decay modes are combined to measure the same strong phase parameters~\cite{Libby:2010nu}, 
but the description below will concentrate only on $\Dz\Dzb$ pairs where both $D$ mesons are decaying to \kpp. 

The model-independent technique can be built based on the observation that 
explicit expressions for the functions $C(\dvar)$ and $S(\dvar)$ are not needed 
to obtain $x_{\pm},y_{\pm}$. 
One can derive a number of independent equations from the expressions (\ref{eq:pdd_dp})
and (\ref{eq:pb_dp}) by integrating both the right and left parts of the equations 
multiplied by certain weight functions $w_n(\DP)$ from a family of functions 
indexed by $1\leq n\leq M$. Equation (\ref{eq:pdd_dp}) then becomes
\begin{equation}
  p_{DD,mn} \equiv \!\!\int\limits_{\DP_1,\DP_2}\!\!w_m(\dvar_1)w_n(\dvar_2)p_{DD}(\dvar_1, \dvar_2)d\dvar_1 d\dvar_2 = 
  h_{DD}\left\{ p_m \bar{p}_n + \bar{p}_m p_n - 2[C_m C_n + S_m S_n] \right\}, 
  \label{eq:pdd_w}
\end{equation}
while Eqs.~(\ref{eq:pb_dp}) become
\begin{equation}
  \begin{split}
  \bar{p}_{\B,n}\equiv\int\limits_{\DP}w_n(\dvar) \bar{p}_{\B}(\dvar) d\dvar & = 
    \bar{h}_{\B}\left\{ \bar{p}_n + r_B^2 p_n + 2[x_+C_n - y_+S_n] \right\}, \\
  p_{\B,n}\equiv\int\limits_{\DP}w_n(\dvar) p_{B}(\dvar) d\dvar & = 
    h_{\B}\left\{ p_n + r_B^2 \bar{p}_n + 2[x_-C_n + y_-S_n] \right\}, 
  \end{split}
  \label{eq:pb_w}
\end{equation}
where 
\begin{equation}
  p_n = \int\limits_{\DP}w_n(\dvar) p_D(\dvar) d\dvar\;\;\;\;
  \bar{p}_n = \int\limits_{\DP}w_n(\dvar) \bar{p}_D(\dvar) d\dvar, 
\end{equation}
and 
\begin{equation}
  C_n = \int\limits_{\DP}w_n(\dvar) C(\dvar) d\dvar, \;\;\;\;
  S_n = \int\limits_{\DP}w_n(\dvar) S(\dvar) d\dvar.
  \label{eq:csn}
\end{equation}
The integration in Eqs.~(\ref{eq:pb_w}--\ref{eq:csn}) is performed over the entire 
Dalitz plot $\DP$ of the $D$ decay, while for Eq.~(\ref{eq:pdd_w}) double integral 
is performed over the Dalitz plots $\DP_1$ and $\DP_2$ of two decaying $D$ mesons. 
Unlike in the binned formalism described in Refs.~\cite{Bondar:2005ki,Bondar:2008hh}, 
here the terms proportional to $|A(\dvar)|\cdot|\overline{A}(\dvar)|$ are not factored out, 
thus capital letters are used to distinguish the expressions of Eq.~(\ref{eq:csn}) 
from $c_i$ and $s_i$ coefficients commonly used in the binned formalism. 

The values of weighted integrals for the flavour-specific $D$ sample ($p_n$ and $\bar{p}_n$), 
$B$ sample ($\bar{p}_{\B,n}$ and $p_{\B,n}$) and correlated \ddb sample ($p_{DD,mn}$)
can be obtained directly from each of the corresponding scattered data samples 
by replacing the integrals with sums over individual observed events. 
The values of the weighted integrals for the phase terms $C_n$ and $S_n$ are considered 
as free parameters constrained by Eq.~(\ref{eq:pdd_w}). This allows 
the values of $x_{\pm}$ and $y_{\pm}$ to be obtained by solving the system of equations (\ref{eq:pdd_w}) 
and (\ref{eq:pb_w}). 

The family of weight functions $w_n$ can be chosen arbitrarily, but the performance of 
the method with a limited data sample will depend on this choice. The binned model-independent 
approach is a particular case of the considered formalism where the weight functions are 
of the form
\begin{equation}
  w_n(\dvar) = \left\{ \begin{array}{l} 
                  1 \mbox{ if }\dvar\in\DP_n\\ 
                  0 \mbox{ otherwise}
                \end{array}\right. .
\end{equation}
Here $\DP_n$ are non-overlapping regions of the Dalitz plot
which define the bins. 

To reach optimal statistical sensitivity, the binning has to be chosen in such a way as to 
maximise the interference term in Eq.~(\ref{eq:pb_w}). A good approximation 
to the optimum is known to be the binning based on the strong phase difference between the 
favoured and suppressed $D$ decay amplitudes~\cite{Bondar:2008hh}. Specifically, 
if one defines the phase difference $\Phi(m^2_+, m^2_-)$ as 
\begin{equation}
  \Phi(m^2_+, m^2_-) = \arg A^{\rm (model)}_D(m^2_+, m^2_-)-\arg A^{\rm (model)}_D(m^2_-, m^2_+), 
  \label{eq:phi}
\end{equation}
then the bin $\DP_n$ ($1\leq n\leq M$) is the region of the phase space which satisfies
\begin{equation}
  2\pi(n-1/2)/M < \Phi(m^2_+, m^2_-) < 2\pi(n+1/2)/M; \;\; m^2_+<m^2_-. 
\end{equation}
The bins in the region with $m^2_+ > m^2_-$ are defined symmetrically with respect to 
exchange $m^2_+ \leftrightarrow m^2_-$ and have indices $n<0$. Here $A^{\rm (model)}_{D}(\dvar)$
is an amplitude model that ideally should approach the true amplitude $A_D(\dvar)$
to reach optimal statistical precision, but does not need to match it exactly to 
provide an unbiased measurement. 

The following section shows how to construct an unbinned model-independent 
formalism using a model-based phase difference function $\Phi(m^2_+, m^2_-)$ 
which will be a generalisation of the technique with phase-difference binning. 
For reasons which will become obvious, this approach will not be optimal from 
the point of view of statistical uncertainty, and will serve solely as a demonstration. 
Subsequently, a more optimal approach based on a similar construction will be presented.

\section{Unbinned technique using Fourier series expansion of phase difference}

\label{sec:fourier-non-split}

Let $\Phi(\dvar)\equiv \Phi(m^2_+, m^2_-)$ be the function defined by Eq.~(\ref{eq:phi})
that maps two-dimensional Dalitz plot coordinates $\dvar$ to the one-dimensional 
space represented by a phase difference $\phi$ between the \Dzb and \Dz amplitudes
at the same Dalitz plot point. 
One can now define probability densities as functions of $\phi = \Phi(\dvar)$. 
The density of the flavour-specific $D$ decay becomes 
\begin{equation}
  p_D(\phi) = \!\!\int\limits_{\Phi(\dvar)=\phi}\!\!p_D(\dvar) d\dvar. 
  \label{eq:pd}
\end{equation}
From the experimentalist's point of view, this function is the probability density (PDF) of 
the $\Phi(\dvar)$ value for a sample of flavour-specific \dkpp decays, 
and is a continuous generalisation of the number of 
events $K_n$ that enters the $n^{\rm th}$ bin in the approach with binning based on 
equal phase difference~\cite{Bondar:2008hh}. Following Eqs.~(\ref{eq:pdbar_dp}) and (\ref{eq:phi}), 
the density for the \CP-conjugate decay is 
\begin{equation}
  \bar{p}_{D}(\phi)=p_D(-\phi).
  \label{eq:pdbar}
\end{equation} 
After a similar mapping is applied to the correlated densities of the two \dkpp Dalitz plots
of the \ddb sample (\ref{eq:pdd_dp}), the following PDF of the variables 
$\phi_1 = \Phi(\dvar_1)$ and $\phi_2 = \Phi(\dvar_2)$ is obtained: 
\begin{equation}
  p_{DD}(\phi_1, \phi_2)=h_{DD}\left\{p_D(\phi_1) \bar{p}_D(\phi_2) + \bar{p}_D(\phi_1) p_D(\phi_2)
                        -2\left[ C(\phi_1)C(\phi_2) + S(\phi_1)S(\phi_2)\right]\right\},
  \label{eq:pdd}
\end{equation}
where
\begin{equation}
  C(\phi) = \!\!\int\limits_{\Phi(\dvar)=\phi}\!\!C(\dvar)d\dvar, \;\;\;
  S(\phi) = \!\!\int\limits_{\Phi(\dvar)=\phi}\!\!S(\dvar)d\dvar, \;\;\;
  \label{eq:cs_phi}
\end{equation}
From the definitions (\ref{eq:cs_phi}) and (\ref{eq:phi}) it follows that 
$C(\phi)$ is an even function, while $S(\phi)$ is odd:
\begin{equation}
  C(-\phi)=C(\phi),\;\;\;S(-\phi) = -S(\phi). 
\end{equation}

Switching to the phase-difference representation for the \bpmdk densities (\ref{eq:pb_dp}), 
one obtains
\begin{equation}
  \begin{split}
  \bar{p}_{\B}(\phi) = \bar{h}_{\B}\left\{\bar{p}_D(\phi) + r_B^2 p_D(\phi) + 2[x_+ C(\phi) - y_+ S(\phi)]\right\},\\
  p_{\B}(\phi) = h_{\B}\left\{p_D(\phi) + r_B^2 \bar{p}_D(\phi) + 2[x_- C(\phi) + y_- S(\phi)]\right\}.
  \end{split}
  \label{eq:pb}
\end{equation}

The next step is to choose the family of weight functions to construct a system of equations 
which allow the determination of $x_{\pm}$ and $y_{\pm}$ from Eqs.~(\ref{eq:pdd}) and (\ref{eq:pb}). 
Since the densities as functions of $\phi$ are periodic by construction, it appears that the natural 
choice is to use Fourier expansion of the functions of the phase difference, {\it i.e.} use weight
functions of the form $\cos(n\phi)$ and $\sin(n\phi)$, where $n$ is an integer number. The unknowns 
$x_{\pm}$ and $y_{\pm}$ will then enter the system of equations which relates the 
coefficients of the Fourier expansions of the $p_D$, $p_{DD}$, $\bar{p}_{\B}$ and $p_{\B}$ densities. 

Specifically, the functions $p_D(\phi)$, $C(\phi)$ and $S(\phi)$ can be represented as 
\begin{equation}
  p_D(\phi) = \frac{a^D_0}{2} + \sum\limits_{n=1}^{M} [ a^D_n \cos(n\phi) + b^D_n \sin(n\phi) ], 
\end{equation}
\begin{equation}
  \bar{p}_D(\phi) = \frac{a^D_0}{2} + \sum\limits_{n=1}^{M} [ a^D_n \cos(n\phi) - b^D_n \sin(n\phi) ], 
\end{equation}
\begin{equation}
  C(\phi) = \frac{a^C_0}{2} + \sum\limits_{n=1}^{M} a^C_n \cos(n\phi), 
\end{equation}
\begin{equation}
  S(\phi) = \sum\limits_{n=1}^{M} b^S_n \sin(n\phi), 
\end{equation}
keeping in mind that $C(\phi)$ is even and $S(\phi)$ is odd. 
The two-dimensional density $p_{DD}$ is represented by the four sets of Fourier coefficients
$a^{DD}_{nm}$, $b^{DD}_{nm}$, $c^{DD}_{nm}$, and $d^{DD}_{nm}$, defined as 
\begin{equation}
  \begin{split}
  p_{DD}(\phi_1, \phi_2) = & \frac{a^{DD}_{00}}{4} + 
    \sum\limits_{m=1}^{M}\frac{a^{DD}_{m0}}{2}\cos(m\phi_1) + 
    \sum\limits_{n=1}^{M}\frac{a^{DD}_{0n}}{2}\cos(n\phi_2) + \\
  & \sum\limits_{n=1}^{M}\frac{b^{DD}_{0n}}{2}\sin(n\phi_2) + 
    \sum\limits_{m=1}^{M}\frac{c^{DD}_{m0}}{2}\sin(m\phi_1) + \\
  & \sum\limits_{m,n=1}^{M}[ 
           a^{DD}_{mn}\cos(m\phi_1)\cos(n\phi_2) + 
           b^{DD}_{mn}\cos(m\phi_1)\sin(n\phi_2) + \\
         & \;\;\;\;\;\;\;\;\;\; c^{DD}_{mn}\sin(m\phi_1)\cos(n\phi_2) + 
           d^{DD}_{mn}\sin(m\phi_1)\sin(n\phi_2)
         ]. 
  \end{split}
  \label{eq:pdd_fourier}
\end{equation}
Strictly speaking, the equations above are exact only in the limit $M\to \infty$, 
however, in practice one has to truncate the Fourier series at a certain finite $M$.

For $p_D(\phi)$, the values of the Fourier coefficients can be calculated directly from 
scattered data $\phi^{(i)}$, $i=1\ldots N_D$: 
\begin{equation}
  a^D_n = \frac{1}{\pi}\sum\limits_{i=1}^{N_D}\cos(n\phi^{(i)}), \;\;
  b^D_n = \frac{1}{\pi}\sum\limits_{i=1}^{N_D}\sin(n\phi^{(i)}),
  \label{eq:ab}
\end{equation}
where $N_D$ is the number of events in the data sample and $\phi^{(i)}=\Phi(\dvar^{(i)})$
are the calculated phase difference values for the data sample entries $\dvar^{(i)}$. 
Similarly, the coefficients of the Fourier expansion for the correlated \ddb sample 
can be calculated from the 2D scattered data $\phi^{(i)}_1=\Phi(\dvar^{(i)}_1), 
\phi^{(i)}_2=\Phi(\dvar^{(i)}_2)$, $i=1\ldots N_{DD}$ as 
\begin{equation}
  \begin{split}
    a^{DD}_{mn} = \frac{1}{\pi}\sum\limits_{i=1}^{N_{DD}}\cos(m\phi_1^{(i)})\cos(n\phi_2^{(i)}), \;\; &
    b^{DD}_{mn} = \frac{1}{\pi}\sum\limits_{i=1}^{N_{DD}}\cos(m\phi_1^{(i)})\sin(n\phi_2^{(i)}), \\
    c^{DD}_{mn} = \frac{1}{\pi}\sum\limits_{i=1}^{N_{DD}}\sin(m\phi_1^{(i)})\cos(n\phi_2^{(i)}), \;\; &
    d^{DD}_{mn} = \frac{1}{\pi}\sum\limits_{i=1}^{N_{DD}}\sin(m\phi_1^{(i)})\sin(n\phi_2^{(i)}). \\
  \end{split}
  \label{eq:abcd}
\end{equation}
On the other hand, from Eq.~(\ref{eq:pdd}) one can obtain a set of relations between the 
Fourier coefficients for flavour-specific and \ddb densities: 
\begin{equation}
  \begin{split}
  a^{DD}_{mn} & = 2 h_{DD}\left(a^D_m a^D_n - a^C_m a^C_n\right), \\
  b^{DD}_{mn} & = c^{DD}_{mn} = 0, \\
  d^{DD}_{mn} & = -2 h_{DD}\left(b^D_m b^D_n + b^S_m b^S_n\right). \\
  \end{split}
  \label{eq:a}
\end{equation}
The expressions (\ref{eq:a}) 
can be used to obtain the unknown coefficients $a^C_n$ and $b^S_n$ from the 
known values of $(a,b)^{D}_n$ and $(a,b,c,d)^{DD}_{mn}$. 
The system of equations (\ref{eq:a}) is solvable for any $M\geq 1$ (there are $2M^2+M+1$ independent 
equations and $2M+2$ unknown parameters). In practice, since the system of equations is 
overconstrained for $M>1$, it should be solved using a maximum likelihood fit, 
which will also provide estimate of the covariance matrix. 

A maximum likelihood fit needs uncertainties for the coefficients that enter the equations. 
These can be calculated analytically by applying a Poisson bootstrapping technique~\cite{Oza01onlinebagging}. Each term entering the sum in Eq.~(\ref{eq:ab}) or (\ref{eq:abcd}) is multiplied by a random number which follows the Poisson 
distribution with unit mean value. 
The variances for the sums can then be obtained assuming they have a Gaussian 
distribution (which is a valid assumption for large $N_D$): 
\begin{equation}
  \sigma^2(a^D_n) = \frac{1}{\pi}\sum\limits_{i=1}^{N_D}\cos^2(n\phi^{(i)}), \;\;
  \sigma^2(b^D_n) = \frac{1}{\pi}\sum\limits_{i=1}^{N_D}\sin^2(n\phi^{(i)}),
  \label{eq:sigma_ab}
\end{equation}
and 
\begin{equation}
  \begin{split}
    \sigma^2(a^{DD}_{mn}) = \frac{1}{\pi}\sum\limits_{i=1}^{N_{DD}}\cos^2(m\phi_1^{(i)})\cos^2(n\phi_2^{(i)}), \;\; &
    \sigma^2(b^{DD}_{mn}) = \frac{1}{\pi}\sum\limits_{i=1}^{N_{DD}}\cos^2(m\phi_1^{(i)})\sin^2(n\phi_2^{(i)}), \\
    \sigma^2(c^{DD}_{mn}) = \frac{1}{\pi}\sum\limits_{i=1}^{N_{DD}}\sin^2(m\phi_1^{(i)})\cos^2(n\phi_2^{(i)}), \;\; &
    \sigma^2(d^{DD}_{mn}) = \frac{1}{\pi}\sum\limits_{i=1}^{N_{DD}}\sin^2(m\phi_1^{(i)})\sin^2(n\phi_2^{(i)}). \\
  \end{split}
  \label{eq:sigma_abcd}
\end{equation}

In addition, unlike in the binned case where the yields in each of the bins are 
statistically independent, the coefficients of the Fourier series are in general 
correlated. The covariance matrix can be 
calculated similarly using Poisson bootstrapping, {\it e.g.} the covariance between the 
$a_n$ and $b_m$ coefficients can be calculated as:
\begin{equation}
  {\rm cov}(a^D_n, b^D_m) = \frac{1}{\pi}\sum\limits_{i=1}^{N_D}\cos(n\phi^{(i)})\sin(m\phi^{(i)}). 
\end{equation} 
Similarly, the expressions for covariances between $a_n$ and $a_m$, $b_n$ and $b_m$, 
or between the coefficients $(a,b,c,d)^{DD}_{mn}$ can be obtained. 

Once the coefficients $(a,b)^{C,S}_n$ are obtained, they can be used to constrain the 
values of $x_{\pm}, y_{\pm}$ (and thus $\gamma$). 
Taking Fourier expansions of the functions $\bar{p}_{\B}(\dvar)$ and $p_{\B}(\dvar)$
\begin{equation}
  \begin{split}
  \bar{p}_{\B}(\phi) & = \frac{\bar{a}^B_0}{2} + \sum\limits_{n=1}^{M} [ \bar{a}^B_n \cos(n\phi) + \bar{b}^B_n \sin(n\phi) ], \\
  p_{\B}(\phi)       & = \frac{      a^B_0}{2} + \sum\limits_{n=1}^{M} [       a^B_n \cos(n\phi) +       b^B_n \sin(n\phi) ], 
  \end{split}
\end{equation}
and plugging them into Eq.~(\ref{eq:pb}), one obtains the following system of equations
\begin{equation}
  \begin{split}
    \bar{a}^B_n & = \bar{h}_B\left[\,\,\,\,(1 + r_B^2) a^D_n  + 2x_+ a^C_n\right], \\
    \bar{b}^B_n & = \bar{h}_B\left[-(1 - r_B^2) b^D_n - 2y_+ b^S_n\right]. \\
          a^B_n & =       h_B\left[\,\,\,\,(1 + r_B^2) a^D_n  + 2x_- a^C_n\right], \\
          b^B_n & =       h_B\left[\,\,\,\,(1 - r_B^2) b^D_n  + 2y_- b^S_n\right]. \\
  \end{split}
  \label{eq:abb}
\end{equation}
which can be solved again using a maximum likelihood fit for any $M\geq 1$, after the extraction of the 
coefficients $(\bar{a},\bar{b})^B_n$ and $(a,b)^B_n$ and their uncertainties and correlations from the \bdk sample
in a similar way. Alternatively, both sets of equations (\ref{eq:a}) and (\ref{eq:abb}) can be 
solved simultaneously using a single combined likelihood. 

\begin{figure}
\centering
\includegraphics[width=0.4\textwidth]{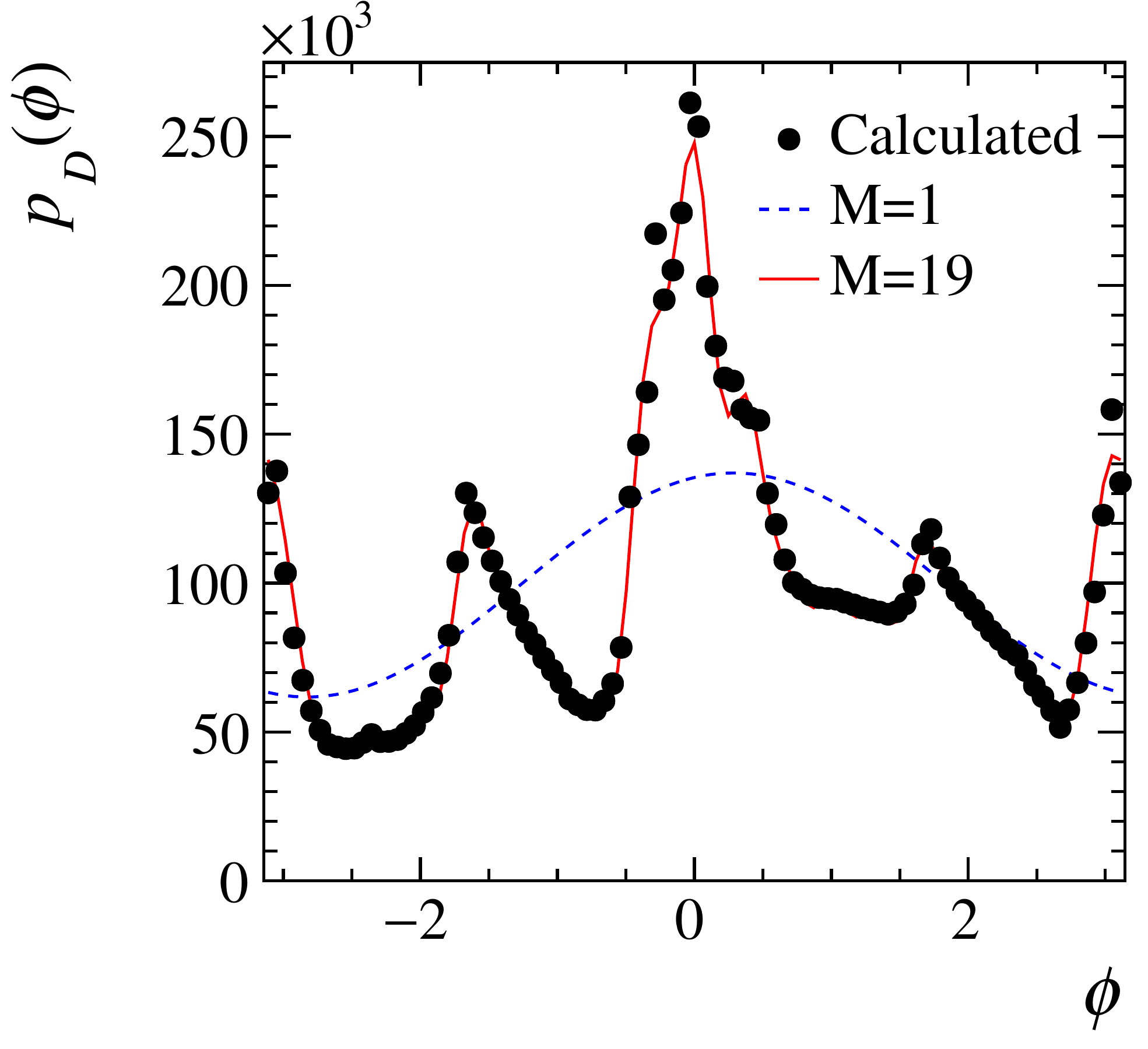}
\put(-130, 140){(a)}

\includegraphics[width=0.4\textwidth]{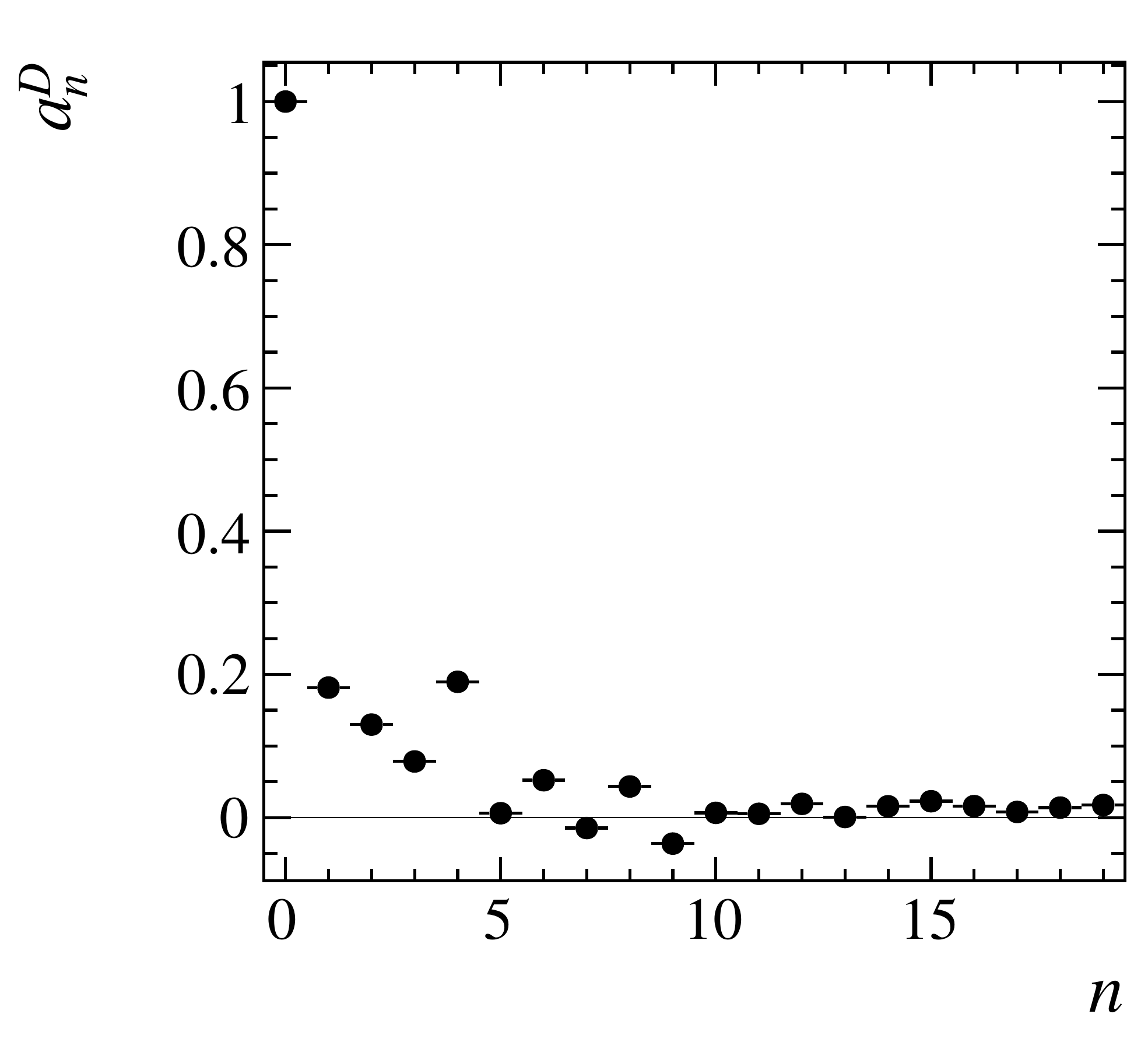}
\put(-28, 140){(b)}
\includegraphics[width=0.4\textwidth]{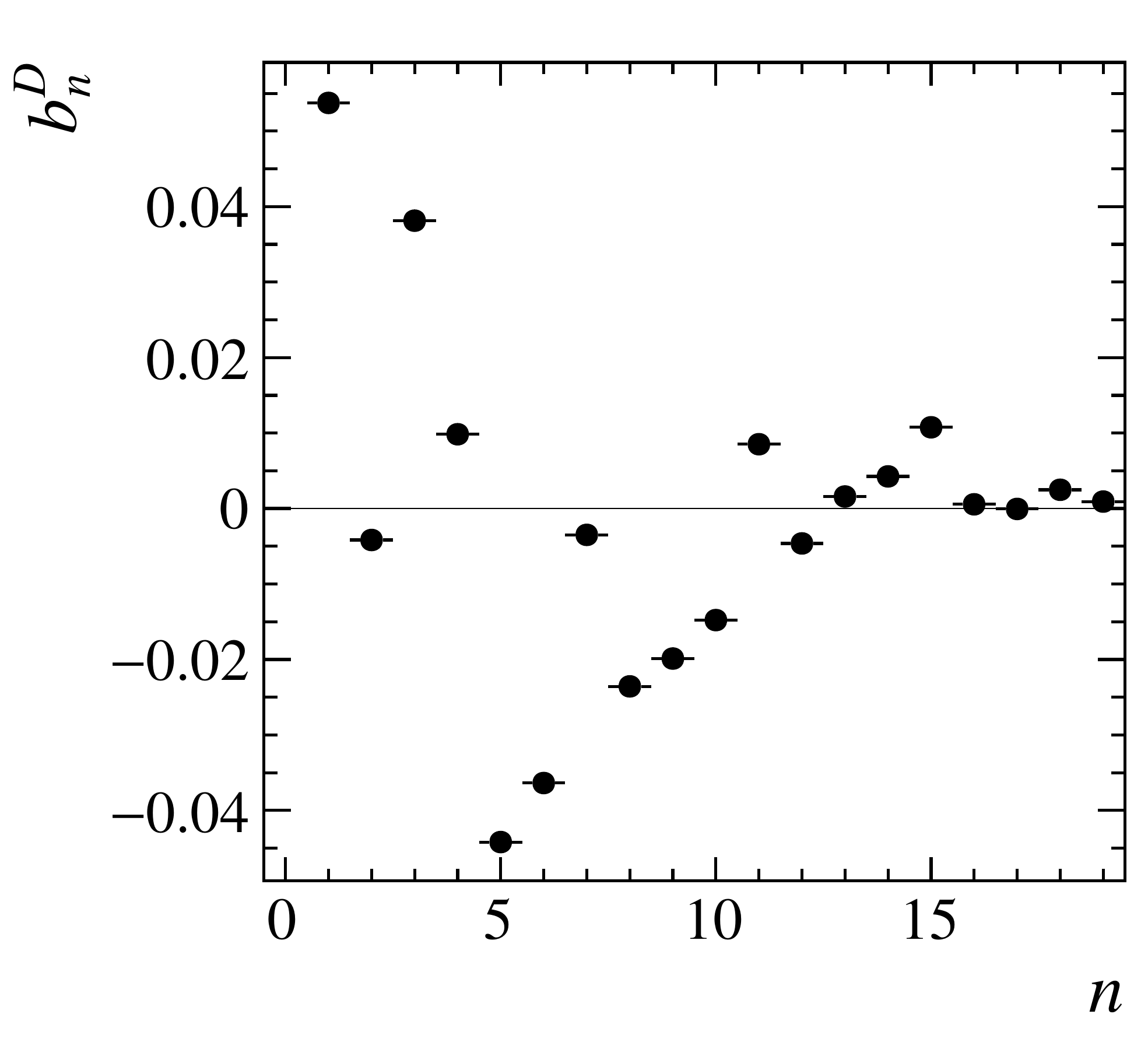}
\put(-28, 140){(c)}

\caption{(a) The function $p_D(\phi)$ according to a model obtained by the Belle 
collaboration~\cite{Poluektov:2010wz}. The points are the histogram of $\phi=\Phi(\dvar)$
values calculated for the generated sample of flavour-specific \dkpp decays, 
solid red line is the result of Fourier expansion with $M=19$, and dashed blue line 
is a single harmonic ($M=1$). (b,c) Fourier series coefficients, calculated from the 
$p_D(\phi)$ density. }
\label{fig:pd}
\end{figure}

\begin{figure}
\centering
\includegraphics[width=0.5\textwidth]{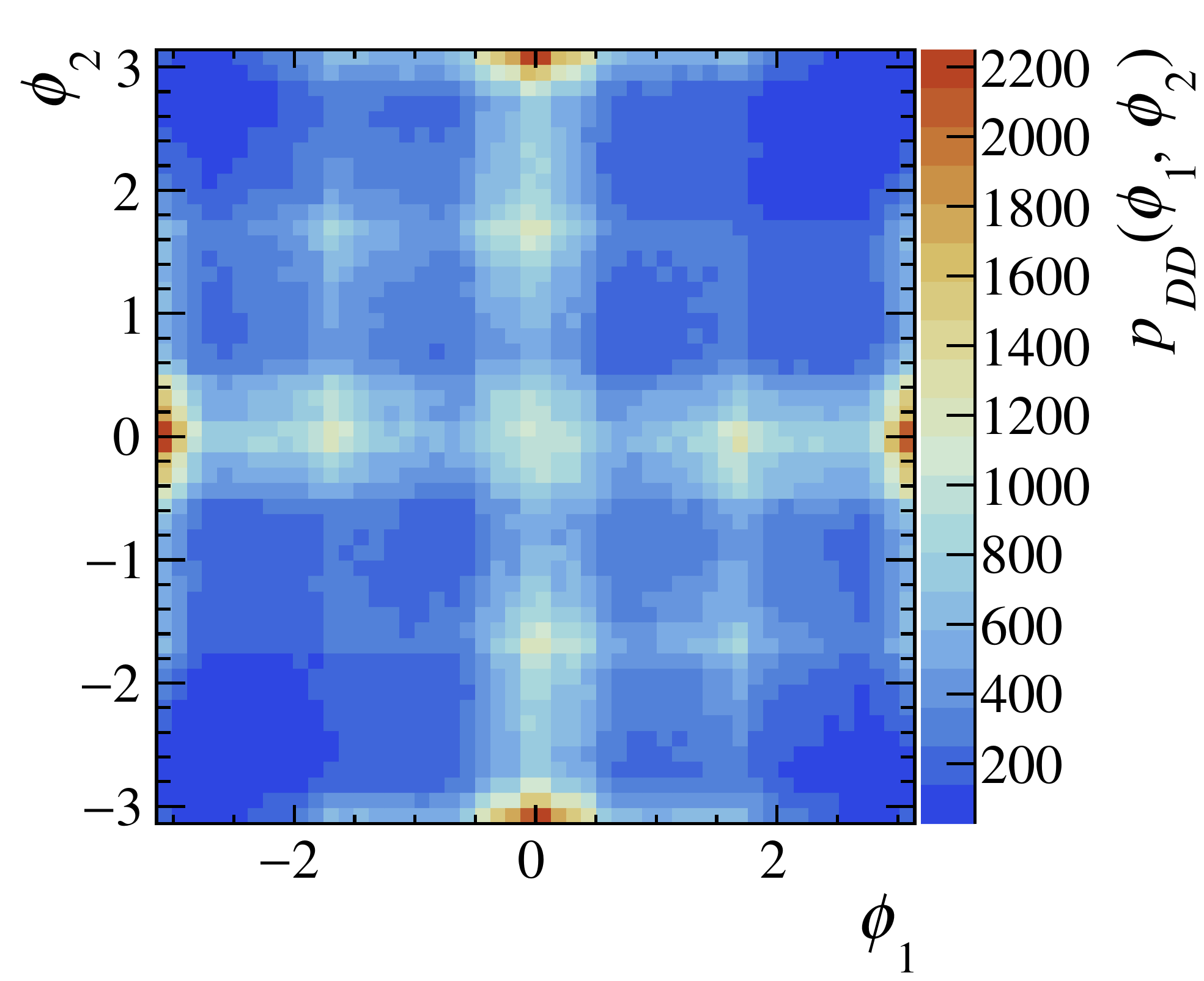}
\put(-80, 160){\colorbox{white}{(a)}}

\includegraphics[width=0.4\textwidth]{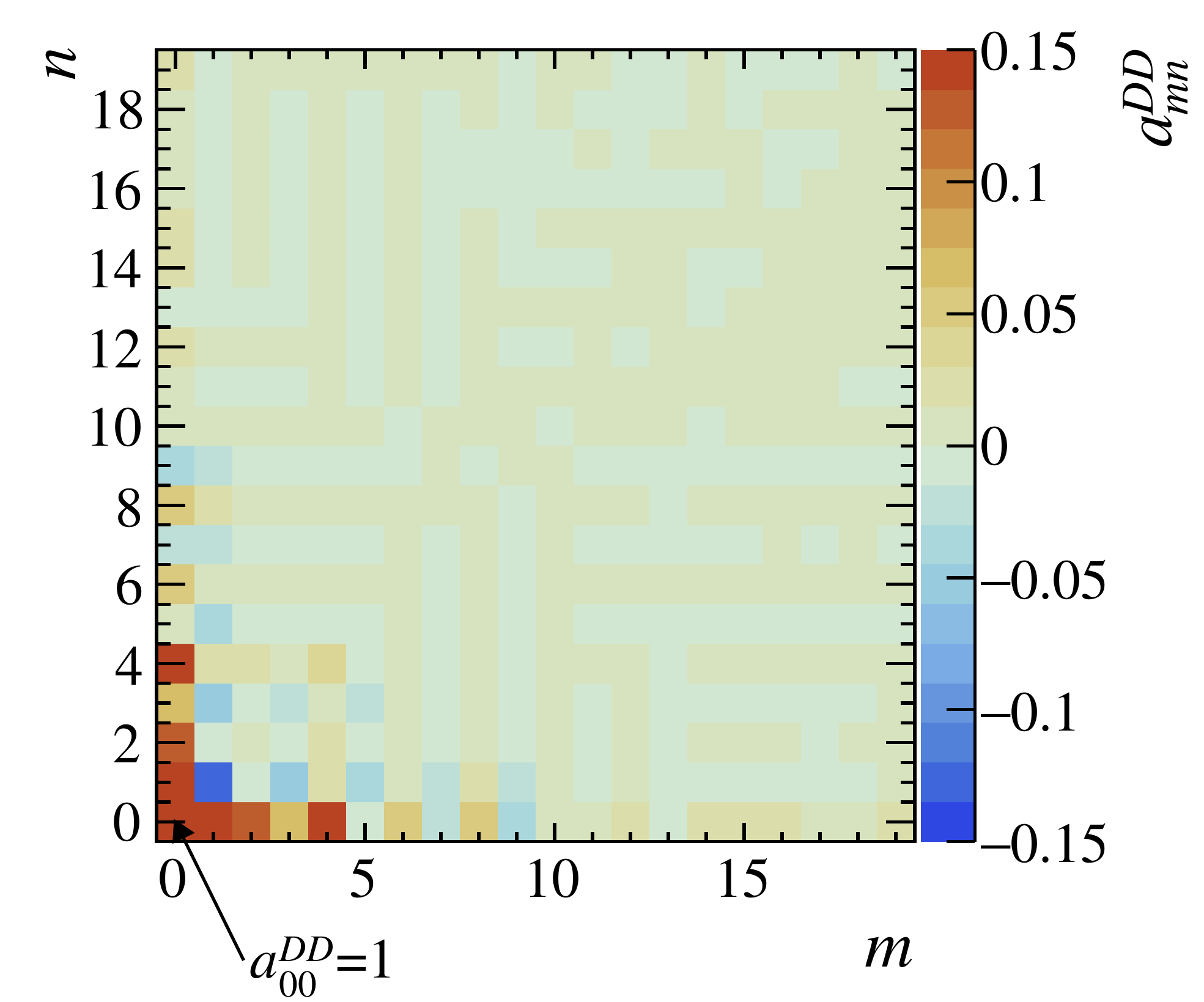}
\put(-70, 125){\colorbox{white}{(b)}}
\includegraphics[width=0.4\textwidth]{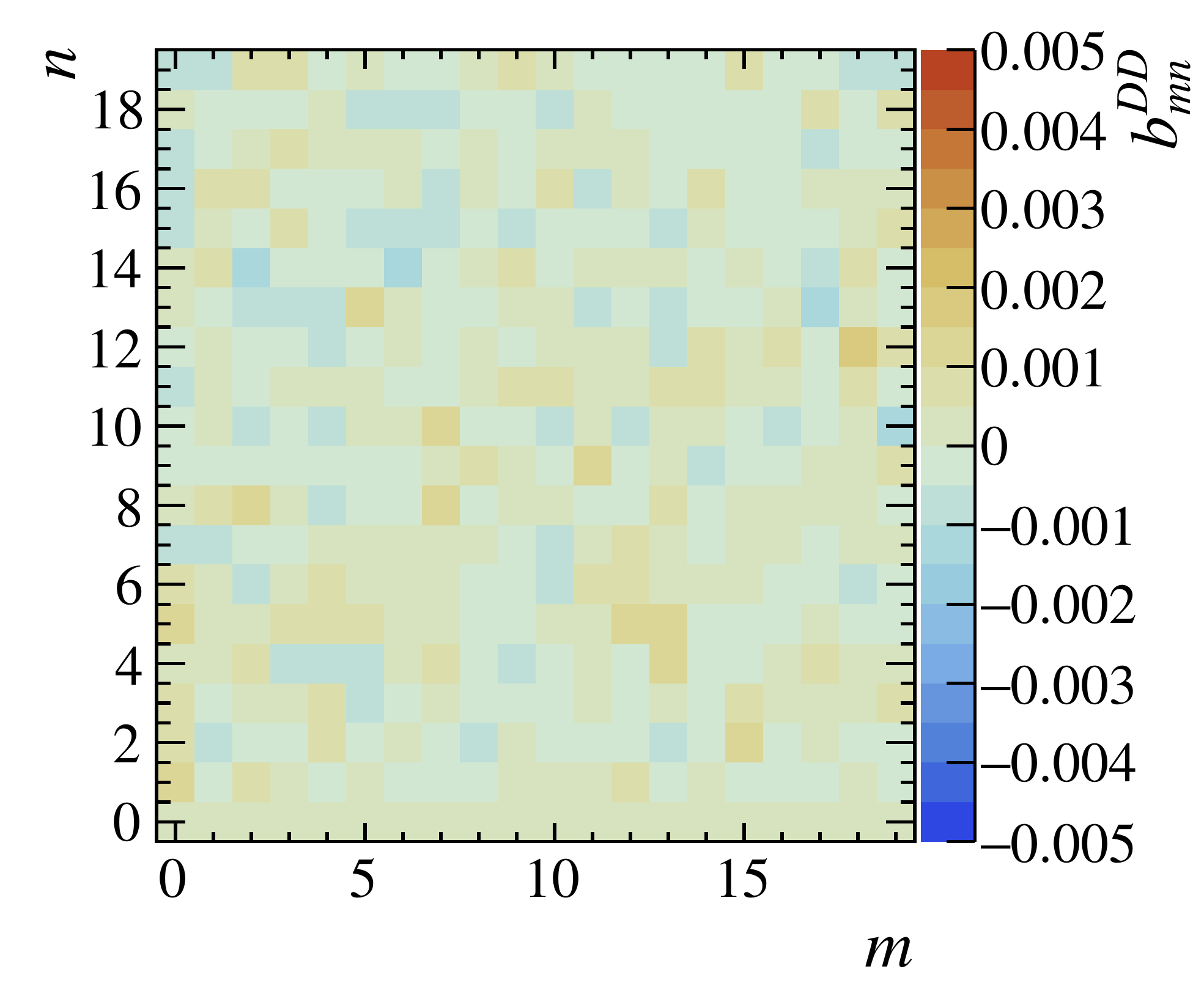}
\put(-70, 125){\colorbox{white}{(c)}}

\includegraphics[width=0.4\textwidth]{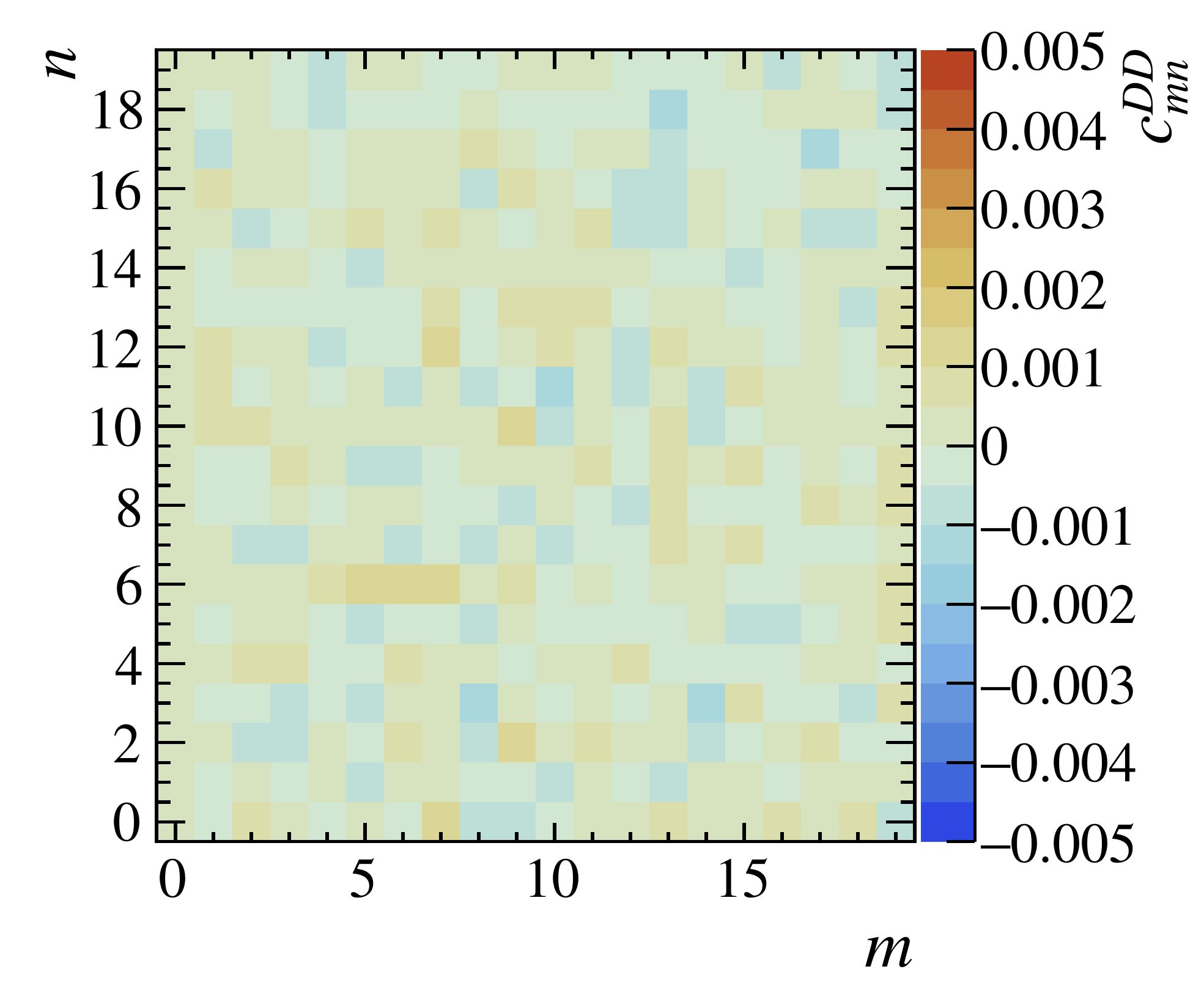}
\put(-70, 125){\colorbox{white}{(d)}}
\includegraphics[width=0.4\textwidth]{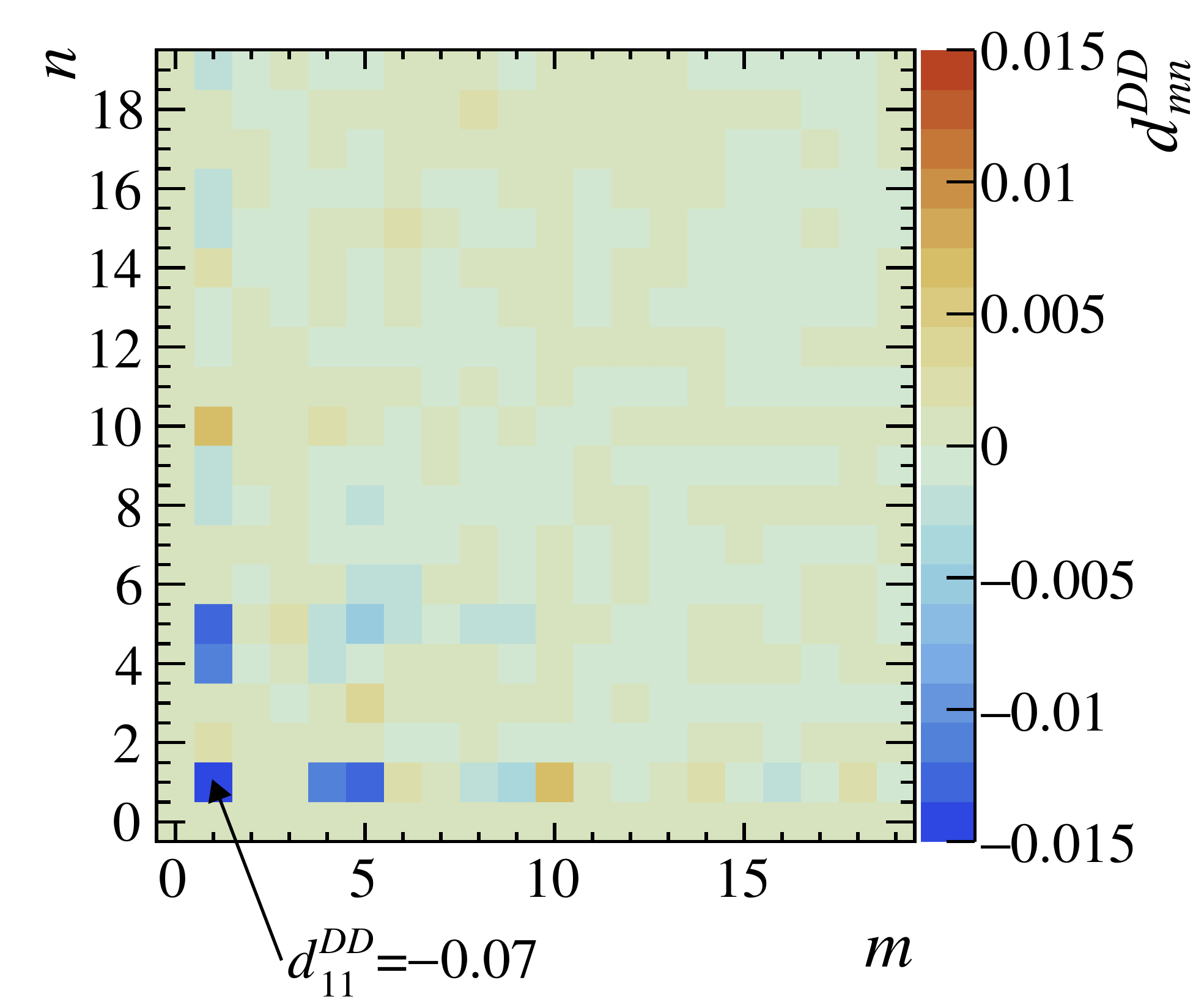}
\put(-70, 125){\colorbox{white}{(e)}}

\caption{(a) The function $p_{DD}(\phi_1, \phi_2)$ according to a model obtained by the Belle
         collaboration~\cite{Poluektov:2010wz}, and (b,c,d,e) its Fourier series coefficients. 
         The ranges of the plots for $a^{DD}_{mn}$ and $d^{DD}_{mn}$ are chosen such that 
         the dominant components $a^{DD}_{00}=1$ and $d^{DD}_{11}=-0.07$ fall out of the range, to 
         make the other components visible. }
\label{fig:pdd}
\end{figure}

\begin{figure}
\centering
\includegraphics[width=0.4\textwidth]{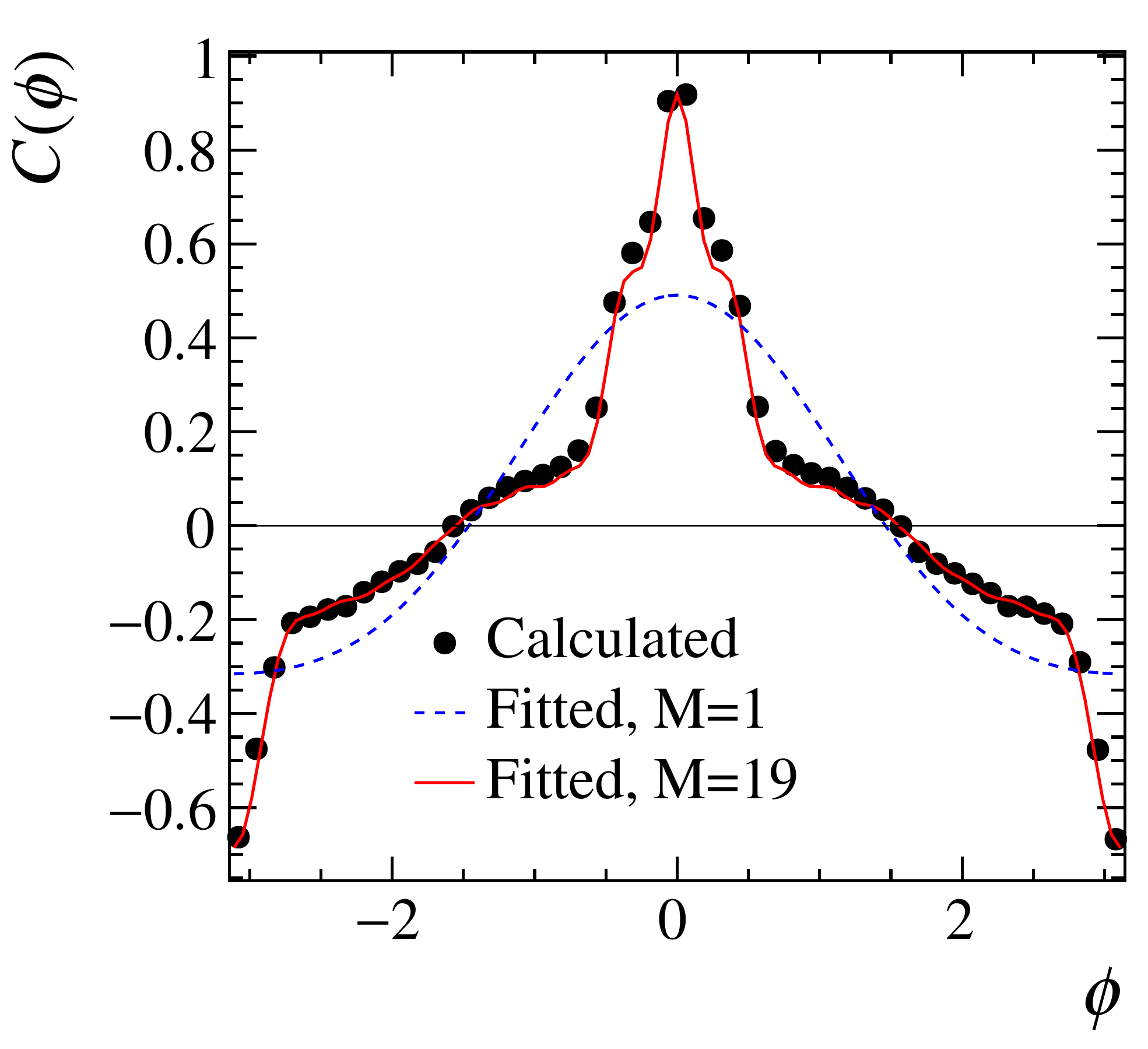}
\put(-30, 140){(a)}
\includegraphics[width=0.4\textwidth]{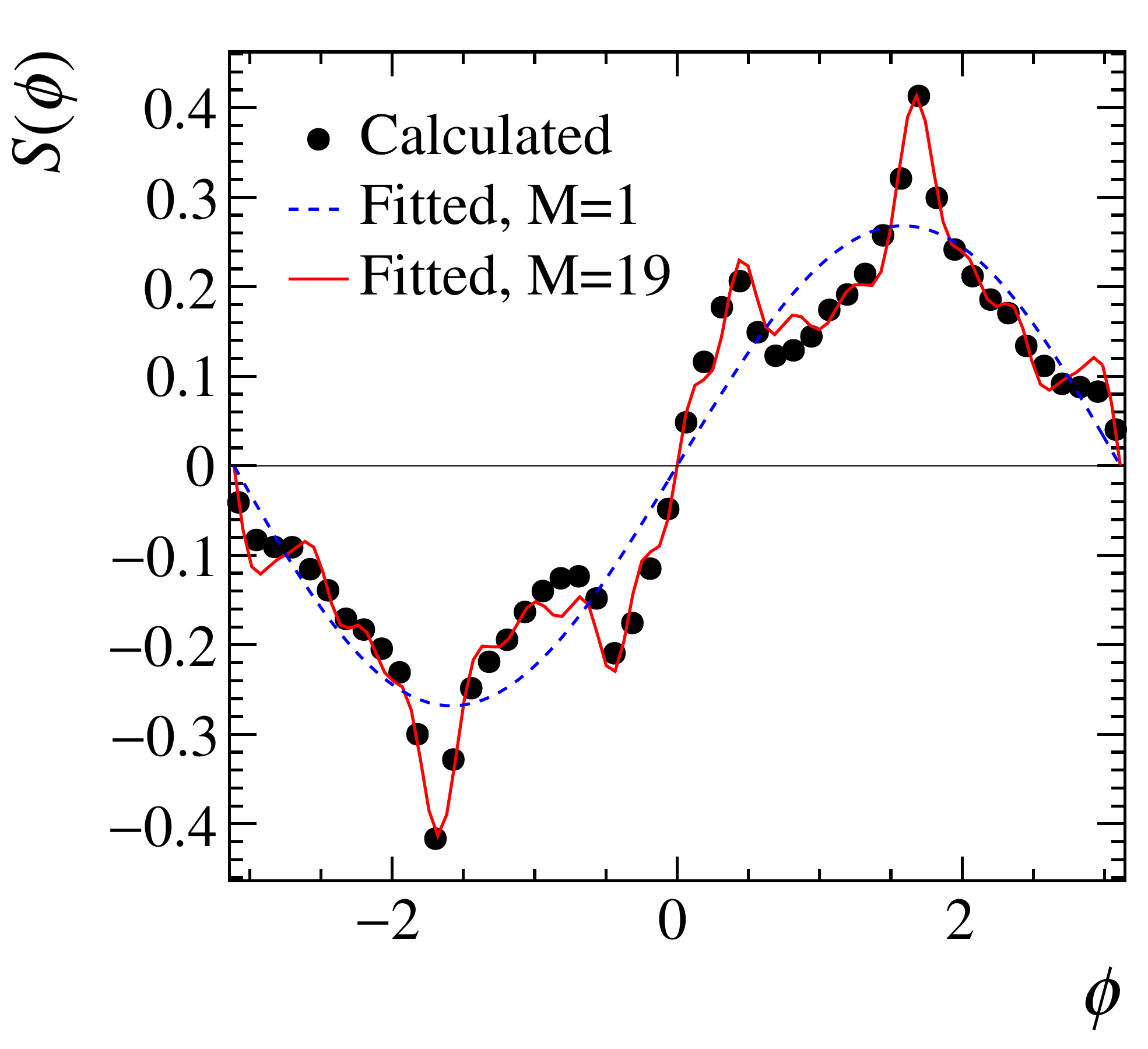}
\put(-135, 40){(b)}

\includegraphics[width=0.4\textwidth]{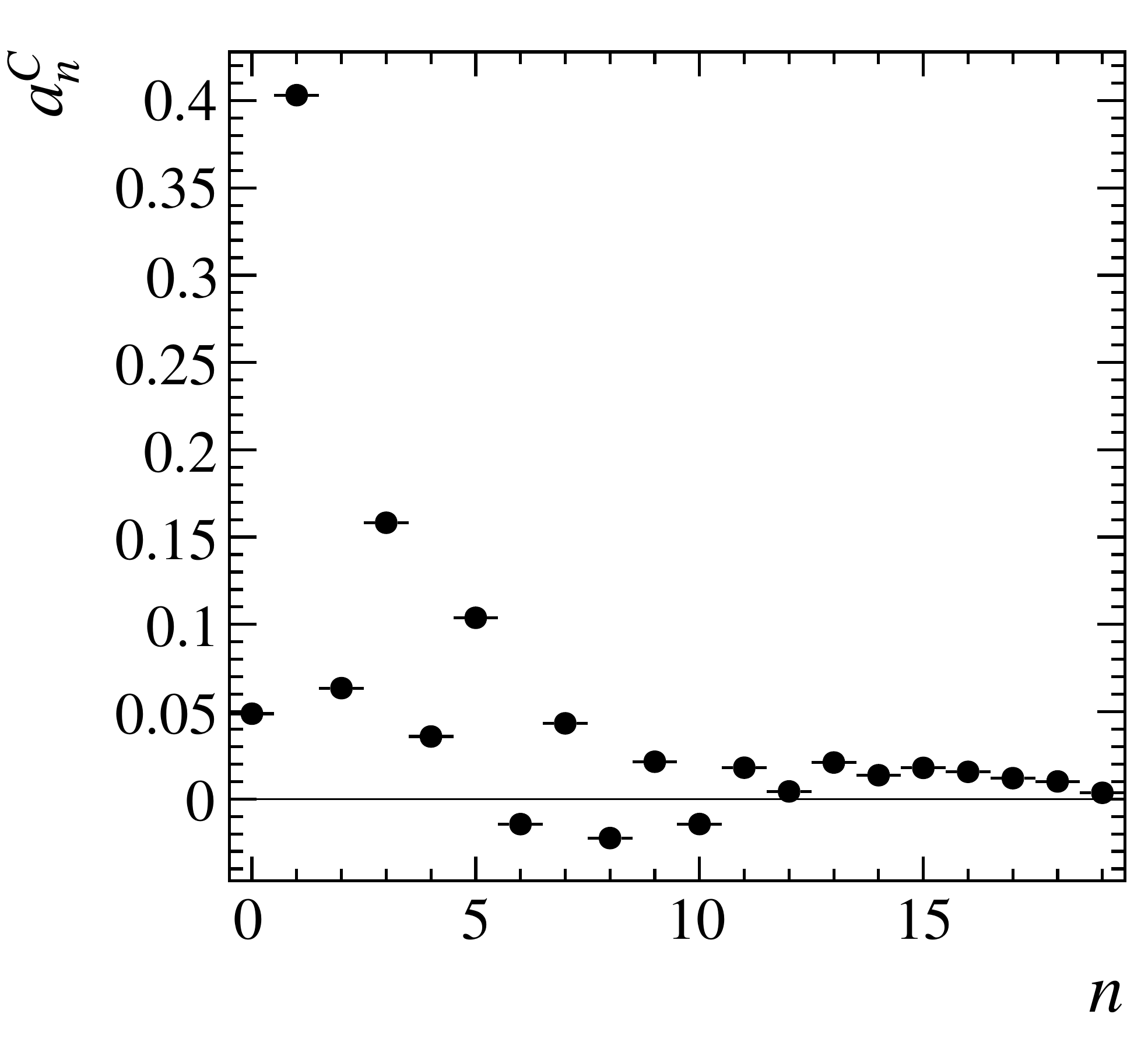}
\put(-30, 140){(c)}
\includegraphics[width=0.4\textwidth]{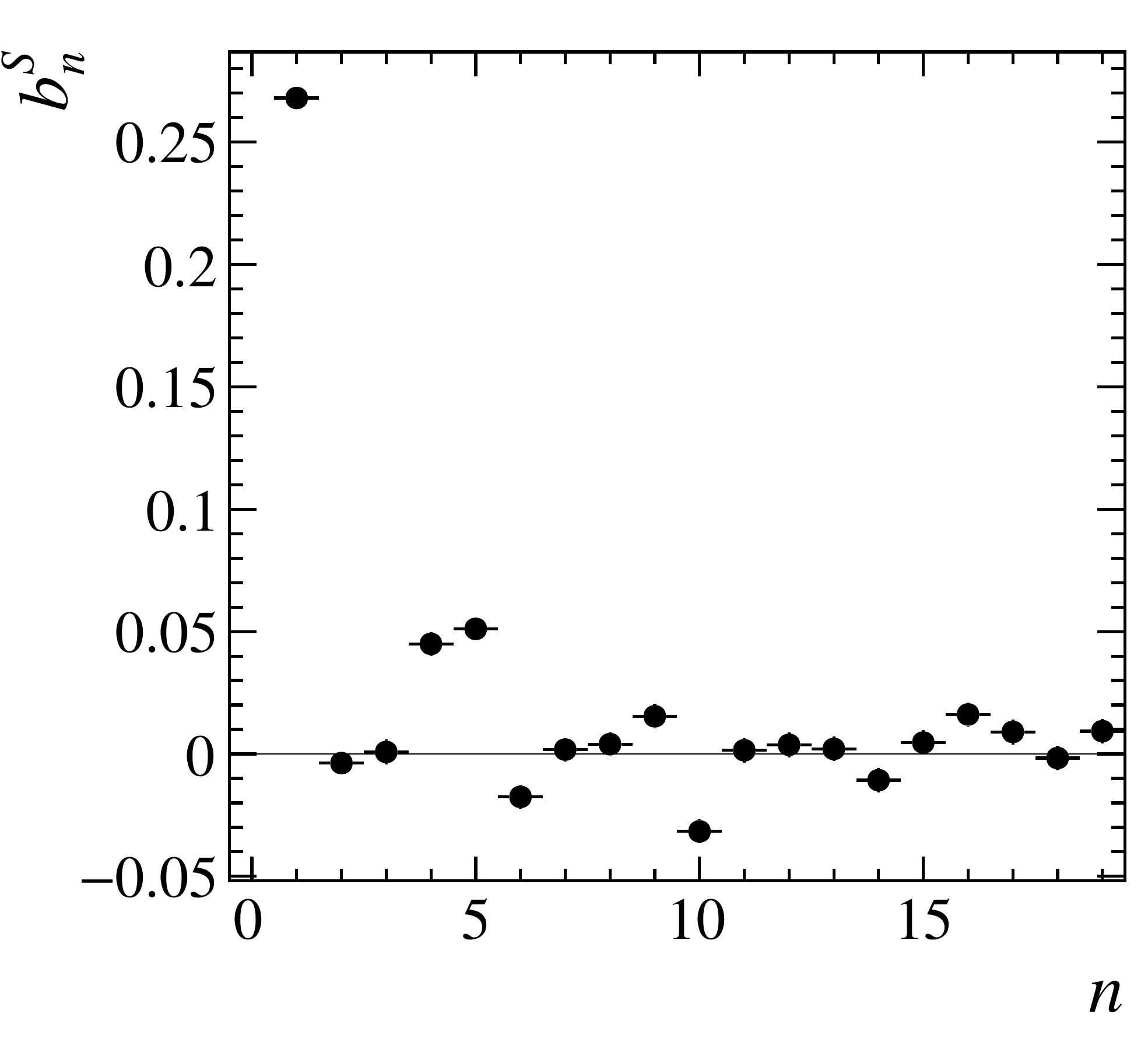}
\put(-30, 140){(d)}
\caption{(a) The functions $C(\phi)$ and (b) $S(\phi)$ according to a model obtained by the Belle
         collaboration~\cite{Poluektov:2010wz}. 
         The points are the functions obtained from the true amplitude $A_D(\dvar)$, 
         solid red line is the function reconstructed from Fourier expansion with $M=19$, 
         and dashed blue line is a single harmonic ($M=1$).
         (c,d) Fourier series coefficients for $C(\phi)$ and $S(\phi)$, respectively. }
\label{fig:cs}
\end{figure}

As an illustration, the functions $p_D(\phi)$, $p_{DD}(\phi_1,\phi_2)$, $C(\phi)$ and $S(\phi)$ 
obtained using the \dkpp amplitude model $A_D(m^2_+, m^2_-)$ from the Belle measurement~\cite{Poluektov:2010wz}
and their respective coefficients of the Fourier expansion are shown in 
Figs~\ref{fig:pd}, \ref{fig:pdd} and \ref{fig:cs}. 
The function $p_D(\phi)$ shown in Fig.~\ref{fig:pd}(a) is obtained by plotting the distribution of 
the function $\phi=\Phi(\dvar)$ (black points) for events generated according to PDF $p(\dvar)$. 
Its Fourier coefficients $a^D_n$ and $b^D_n$ up to $n=19$ are shown 
in Figs~\ref{fig:pd}(b) and (c), respectively. Since the normalisation is arbitrary, the coefficients 
are normalised such that $a^{D}_0=1$. 
The solid red line in Fig.~\ref{fig:pd}(a) shows the result of Fourier expansion 
up to $n=19$, and the dashed blue line shows the first harmonic (expansion up to $n=1$). 
In the $p_{DD}(\phi)$ function that is obtained similarly by plotting the two-dimensional 
distribution of $\phi_1=\Phi(\dvar_1), \phi_2=\Phi(\dvar_2)$ for the correlated Dalitz plot 
points generated according to the $p_{DD}(\dvar_1, \dvar_2)$ density (Fig.~\ref{fig:pdd}), 
only the $a^{DD}_{mn}$ and $d^{DD}_{mn}$ coefficients are non-zero, 
while $b^{DD}_{mn}$ and $c^{DD}_{mn}$ are consistent with zero as expected from Eq.~(\ref{eq:a}). 
The normalisation $a^{DD}_{00}=1$ is used. 

The true functions $C(\phi)$ and $S(\phi)$ can be obtained, on one hand, from the known amplitude $A_D(\dvar)$
by plotting the distribution of the function $\phi=\Phi(\dvar)$ for events generated uniformly across the 
Dalitz plot with event-by-event weights $\sqrt{p_D(\dvar) \bar{p}_D(\dvar)}\cos \Phi(\dvar)$ and 
$\sqrt{p_D(\dvar) \bar{p}_D(\dvar)}\sin \Phi(\dvar)$, respectively. These functions are shown in Fig.~\ref{fig:cs}(a,b)
as black points. On the other hand, the functions can be reconstructed from the spectral coefficients $a^C_n$ and $b^S_n$ obtained 
from Eq.~(\ref{eq:a}). The fitted coefficients $a^{C}_n$ and $b^{S}_n$ are plotted in Figs.~\ref{fig:cs}(c) and (d), 
while the functions $C(\phi)$ and $S(\phi)$ reconstructed from them are shown in Figs.~\ref{fig:cs}(a) and (b) as
solid red lines (from the coefficients up to $n=19$) and dashed blue line (only one harmonic, $n=1$). 
It can be seen from Figs.~\ref{fig:cs}(c,d) that the highest ``power'' of the $C(\phi)$ and $S(\phi)$ spectrum 
is contained in the first harmonic, $n=1$. As a result, as will be seen from further studies with pseudoexperiments, 
limiting $M=1$ is sufficient to reach good sensitivity to $\gamma$. 

\section{Strategy with Fourier expansion on split Dalitz plot}

\label{sec:fourier-split}

The strategy outlined above is the simplest example of the approach using Fourier expansion 
of the phase difference distribution to measure $\gamma$. However, it is clear that this approach 
is not optimal from the point of view of statistical precision. The reason is that one integrates 
over all points of the phase space with the same expected phase difference, 
regardless of the magnitudes of the interfering \dnkpp and \dbkpp amplitudes. 
This effectively reduces the interference term and, as a consequence, 
the sensitivity to the relative phase between the two amplitudes. For similar reasons, 
the ``optimal'' binning scheme was introduced for the binned model-independent approach 
to improve the precision with the equal phase difference binning~\cite{Bondar:2008hh}. 

The simplest way to improve the situation with the technique described here is to split the Dalitz plot into 
regions with comparable ratios between the absolute values of the interfering amplitudes, and to 
perform Fourier expansion in those regions separately. This approach is illustrated below in 
an example with the Dalitz plot split into two regions. 

\begin{figure}
  \centering
  \includegraphics[width=0.5\textwidth]{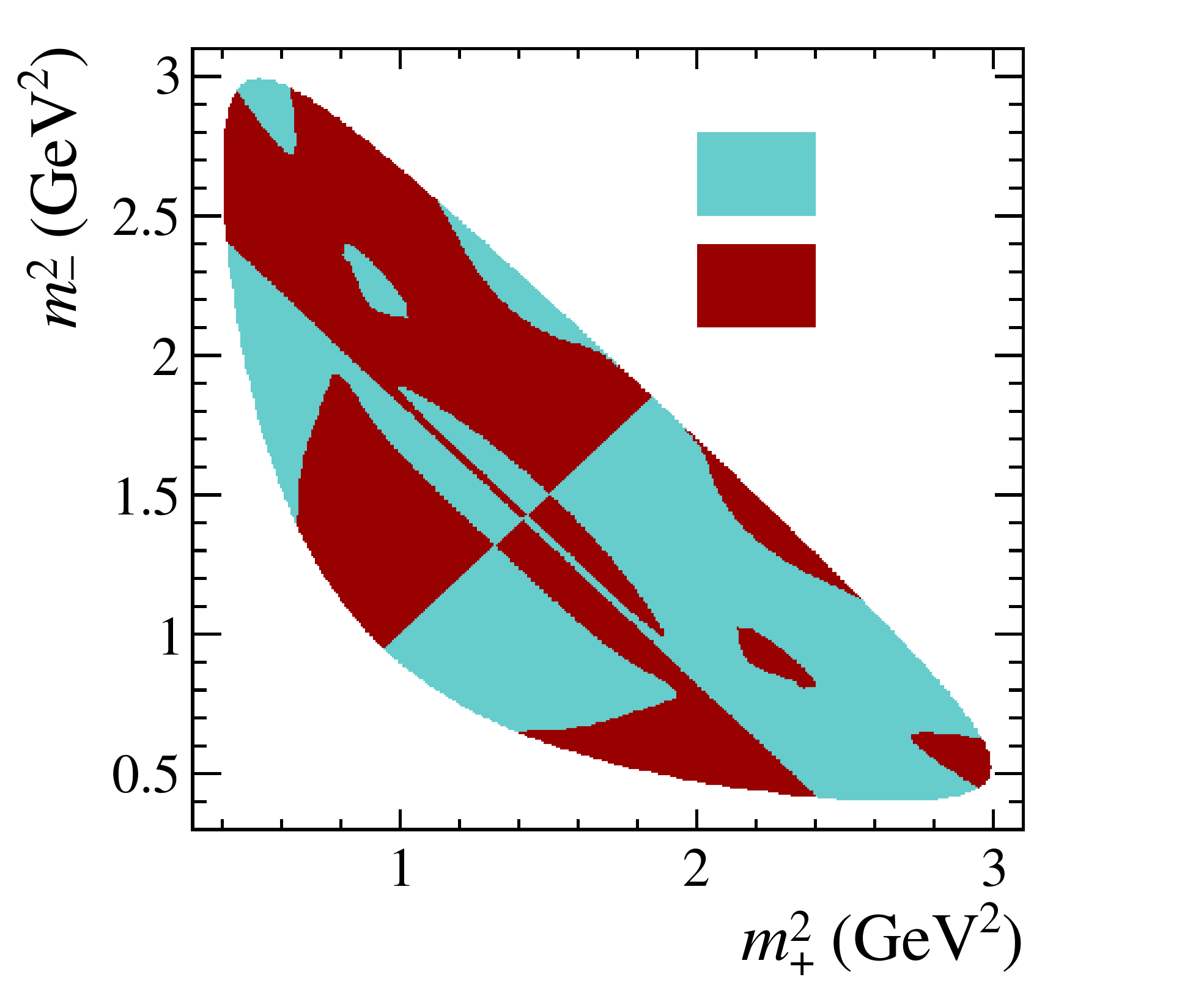}
  \put(-69,150){$\DP^-$}
  \put(-69,128){$\DP^+$}
  \caption{Splitting of \dkpp Dalitz plot into regions $\DP^+$ and $\DP^{-}$. }
  \label{fig:split}
\end{figure}

Two regions of the Dalitz plot are considered: one with $\bar{p}_D(\dvar)>p_{D}(\dvar)$ 
(denoted as region $\DP^+$) and the other with $\bar{p}_D(\dvar)<p_{D}(\dvar)$ (region $\DP^-$). 
These are shown in Fig.~\ref{fig:split} for the same \dkpp amplitude model 
as used in the previous section. 
Clearly, the exchange $m^2_+ \leftrightarrow m^2_-$
transforms $\DP^+$ into $\DP^-$. Now one has to deal with two independent distributions for the 
\dkpp density, $p^+_D$ and $p^-_D$, as functions of the phase difference $\phi$, defined as: 
\begin{equation}
  p^{\pm}_D(\phi) = \!\!\!\!\int\limits_{\Phi(\dvar)=\phi;\;\; \dvar\in \DP^{\pm}}
    \!\!\!\!p_D(\dvar) d\dvar. 
  \label{eq:pd_split}
\end{equation}
The corresponding distributions for the \CP-conjugated decays are 
\begin{equation}
  \bar{p}^{\pm}_D(\phi) = \!\!\!\!\int\limits_{\Phi(\dvar)=\phi;\;\; \dvar\in \DP^{\pm}}
    \!\!\!\!\bar{p}_D(\dvar) d\dvar 
    = p^{\mp}_D(-\phi). 
  \label{eq:pdbar_split}
\end{equation}
It should be stressed that the superscripts ``$+$'' and ``$-$'' denote two Dalitz plot regions 
rather than $B$ meson flavours. 
Throughout this paper, the flavour ($\bquark$ or $\bar{\bquark}$) is consistently 
denoted by the absence or presence 
of a ``bar'' in the corresponding quantities, for example $\bar{p}_{\B}$ and $p_B$, 
except for the subscript for \CP-violating parameters $x_{\pm}, y_{\pm}$ 
which is a commonly used notation. 

With the Dalitz plot split in this way, one needs to define two sets of functions $C(\phi)$ and $S(\phi)$ in 
the two Dalitz plot regions: 
\begin{equation}
  C^{\pm}(\phi) = \!\!\!\!\int\limits_{\Phi(\dvar)=\phi;\;\; \dvar\in \DP^{\pm}}
  \!\!\!\!C(\dvar)d\dvar, \;\;\;
  S^{\pm}(\phi) = \!\!\!\!\int\limits_{\Phi(\dvar)=\phi;\;\; \dvar\in \DP^{\pm}}
  \!\!\!\!S(\dvar)d\dvar. \;\;\;
  \label{eq:cs_split}
\end{equation}
These functions will not be even and odd, as in the previous example, but instead they will satisfy the following properties:  
\begin{equation}
  C^-(\phi) = C^+(-\phi), \;\;\; S^-(\phi) = -S^+(-\phi). 
  \label{eq:cspm_split}
\end{equation}

The two-dimensional density of the $\Dz\Dzb$ sample will be described by a set of four functions
$p_{DD}^{++}$, $p_{DD}^{+-}$, $p_{DD}^{-+}$ and $p_{DD}^{--}$ defined as 
\begin{equation}
  p_{DD}^{s_1 s_2}(\phi_1, \phi_2) = \!\!\!\!\int\limits_{
             \begin{array}{l}\scriptstyle \Phi(\dvar_1)=\phi_1;\;\;  \dvar_1\in \DP^{s_1};  \\ 
                             \scriptstyle \Phi(\dvar_2)=\phi_2;\;\;  \dvar_2\in \DP^{s_2}\end{array}}
                     \!\!\!\! p_{DD}(\dvar_1, \dvar_2) d\dvar_1 d\dvar_2, 
  \label{eq:pdd_split}
\end{equation}
where $s_1, s_2=\{$``$+$'', ``$-$''$\}$. 

The Fourier expansion coefficients $a^{D\pm}_n$ and $b^{D\pm}_n$ for the $D$ decay densities $p^{\pm}_D(\phi)$ are defined as
\begin{equation}
  p^{\pm}_D(\phi) = \frac{a^{D\pm}_0}{2} + \sum\limits_{n=1}^{M} [ a^{D\pm}_n \cos(n\phi) + b^{D\pm}_n \sin(n\phi) ], 
\end{equation}
and similar coefficients for $\bar{p}_D(\phi)$ are denoted as $\bar{a}^{D\pm}_n$ and $\bar{b}^{D\pm}_n$. 
In the case of \CP conservation in $D$ decay, following Eq.~(\ref{eq:pdbar_split}), they are related as 
\begin{equation}
  \bar{a}^{D\pm}_n = a^{D\mp}_n,\;\;\; \bar{b}^{D\pm}_n = -b^{D\mp}_n. 
  \label{eq:abd_sym}
\end{equation}
The Fourier expansion coefficients $(a,b)^{C\pm}_n$ and $(a,b)^{S\pm}_n$ for the 
$C^{\pm}(\phi)$ and $S^{\pm}(\phi)$ functions, respectively, are defined as
\begin{equation}
  C^{\pm}(\phi) = \frac{a^{C\pm}_0}{2} + \sum\limits_{n=1}^{M}[a^{C\pm}_n \cos(n\phi) + b^{C\pm}_n \sin(n\phi)]
\end{equation}
and 
\begin{equation}
  S^{\pm}(\phi) = \frac{a^{S\pm}_0}{2} + \sum\limits_{n=1}^{M}[a^{S\pm}_n \cos(n\phi) + b^{S\pm}_n \sin(n\phi)], 
\end{equation}
and are related as 
\begin{equation}
  a^{C-}_n = a^{C+}_n,\;\;\; b^{C-}_n=-b^{C+}_n,\;\;\;a^{S-}_n=-a^{S+}_n,\;\;\;b^{S-}_n=b^{S+}_n. 
  \label{eq:abcs_sym}
\end{equation}

The relations between the coefficients of the Fourier expansion of the \ddb and flavour 
$D$ decay densities in that case take the following form: 
\begin{equation}
  \begin{split}
  a^{DD++}_{mn} &= h_{DD}[a^{D+}_m \bar{a}^{D+}_n + \bar{a}^{D+}_m a^{D+}_n - 2(a^{C+}_m a^{C+}_n + a^{S+}_m a^{S+}_n)], \\
  a^{DD-+}_{mn} &= h_{DD}[a^{D+}_m \bar{a}^{D-}_n + \bar{a}^{D+}_m a^{D-}_n - 2(a^{C+}_m a^{C-}_n + a^{S+}_m a^{S-}_n)], \\
  a^{DD+-}_{mn} &= h_{DD}[a^{D-}_m \bar{a}^{D+}_n + \bar{a}^{D-}_m a^{D+}_n - 2(a^{C-}_m a^{C+}_n + a^{S-}_m a^{S+}_n)], \\
  a^{DD--}_{mn} &= h_{DD}[a^{D-}_m \bar{a}^{D-}_n + \bar{a}^{D-}_m a^{D-}_n - 2(a^{C-}_m a^{C-}_n + a^{S-}_m a^{S-}_n)], \\[1mm]
  b^{DD++}_{mn} &= h_{DD}[a^{D+}_m \bar{b}^{D+}_n + \bar{a}^{D+}_m b^{D+}_n - 2(a^{C+}_m b^{C+}_n + a^{S+}_m b^{S+}_n)], \\
  b^{DD-+}_{mn} &= h_{DD}[a^{D+}_m \bar{b}^{D-}_n + \bar{a}^{D+}_m b^{D-}_n - 2(a^{C+}_m b^{C-}_n + a^{S+}_m b^{S-}_n)], \\
  b^{DD+-}_{mn} &= h_{DD}[a^{D-}_m \bar{b}^{D+}_n + \bar{a}^{D-}_m b^{D+}_n - 2(a^{C-}_m b^{C+}_n + a^{S-}_m b^{S+}_n)], \\
  b^{DD--}_{mn} &= h_{DD}[a^{D-}_m \bar{b}^{D-}_n + \bar{a}^{D-}_m b^{D-}_n - 2(a^{C-}_m b^{C-}_n + a^{S-}_m b^{S-}_n)], \\[1mm]
  c^{DD++}_{mn} &= h_{DD}[b^{D+}_m \bar{a}^{D+}_n + \bar{b}^{D+}_m a^{D+}_n - 2(b^{C+}_m a^{C+}_n + b^{S+}_m a^{S+}_n)], \\
  c^{DD-+}_{mn} &= h_{DD}[b^{D+}_m \bar{a}^{D-}_n + \bar{b}^{D+}_m a^{D-}_n - 2(b^{C+}_m a^{C-}_n + b^{S+}_m a^{S-}_n)], \\
  c^{DD+-}_{mn} &= h_{DD}[b^{D-}_m \bar{a}^{D+}_n + \bar{b}^{D-}_m a^{D+}_n - 2(b^{C-}_m a^{C+}_n + b^{S-}_m a^{S+}_n)], \\
  c^{DD--}_{mn} &= h_{DD}[b^{D-}_m \bar{a}^{D-}_n + \bar{b}^{D-}_m a^{D-}_n - 2(b^{C-}_m a^{C-}_n + b^{S-}_m a^{S-}_n)], \\[1mm]
  d^{DD++}_{mn} &= h_{DD}[b^{D+}_m \bar{b}^{D+}_n + \bar{b}^{D+}_m b^{D+}_n - 2(b^{C+}_m b^{C+}_n + b^{S+}_m b^{S+}_n)], \\
  d^{DD-+}_{mn} &= h_{DD}[b^{D+}_m \bar{b}^{D-}_n + \bar{b}^{D+}_m b^{D-}_n - 2(b^{C+}_m b^{C-}_n + b^{S+}_m b^{S-}_n)], \\
  d^{DD+-}_{mn} &= h_{DD}[b^{D-}_m \bar{b}^{D+}_n + \bar{b}^{D-}_m b^{D+}_n - 2(b^{C-}_m b^{C+}_n + b^{S-}_m b^{S+}_n)], \\
  d^{DD--}_{mn} &= h_{DD}[b^{D-}_m \bar{b}^{D-}_n + \bar{b}^{D-}_m b^{D-}_n - 2(b^{C-}_m b^{C-}_n + b^{S-}_m b^{S-}_n)]. \\[1mm]
  \end{split}
  \label{eq:abcd_split}
\end{equation}
where the coefficients $(a,b,c,d)^{DDs_1 s_2}_{mn}$ are defined similarly to those in Eq.~(\ref{eq:pdd_fourier}), 
and the two superscripts $s_1, s_2=\{$``$+$'', ``$-$''$\}$ 
correspond to the superscripts of the $p_{DD}^{s_1 s_2}$ functions (\ref{eq:pdd_split}). 
The coefficients $(\bar{a},\bar{b})^{D\pm}$ and $(a,b)^{(C,S)-}$  
can be substituted by $(a,b)^{D\pm}$ and $(a,b)^{(C,S)+}$, respectively, using relations 
(\ref{eq:abd_sym}) and (\ref{eq:abcs_sym}), reducing the number of free parameters to fit. 
This substitution is, however, not done in Eq.~(\ref{eq:abcd_split}) to emphasise the 
symmetry of the equations. 

Finally, the equations for the densities of the $D$ decay from \bpmdk take the following form in the 
split Dalitz plot case: 
\begin{equation}
  \begin{split}
    \bar{a}^{B+}_n & = \bar{h}_B\left\{\bar{a}^{D+}_n + r_B^2 a^{D+}_n + 2[x_+ a^{C+}_n - y_+ a^{S+}_n]\right\}, \\
    \bar{a}^{B-}_n & = \bar{h}_B\left\{\bar{a}^{D-}_n + r_B^2 a^{D-}_n + 2[x_+ a^{C-}_n + y_+ a^{S-}_n]\right\}, \\
    \bar{b}^{B+}_n & = \bar{h}_B\left\{\bar{b}^{D+}_n + r_B^2 b^{D+}_n + 2[x_+ b^{C+}_n - y_+ b^{S+}_n]\right\}, \\
    \bar{b}^{B-}_n & = \bar{h}_B\left\{\bar{b}^{D-}_n + r_B^2 b^{D-}_n + 2[x_+ b^{C-}_n + y_+ b^{S-}_n]\right\}, \\[1mm]
    a^{B+}_n & = h_B\left\{a^{D+}_n + r_B^2 \bar{a}^{D+}_n + 2[x_- a^{C+}_n + y_- a^{S+}_n]\right\}, \\
    a^{B-}_n & = h_B\left\{a^{D-}_n + r_B^2 \bar{a}^{D-}_n + 2[x_- a^{C-}_n - y_- a^{S-}_n]\right\}, \\
    b^{B+}_n & = h_B\left\{b^{D+}_n + r_B^2 \bar{b}^{D+}_n + 2[x_- b^{C+}_n + y_- b^{S+}_n]\right\}, \\
    b^{B-}_n & = h_B\left\{b^{D-}_n + r_B^2 \bar{b}^{D-}_n + 2[x_- b^{C-}_n - y_- b^{S-}_n]\right\}. \\
  \end{split}
  \label{eq:abb_split}
\end{equation}
The number of unknown $D$ phase parameters in the equations has now increased: there are 
$4M+2$ independent coefficients $(a,b)^{C,S+}_n$ ($0\leq n\leq M$ for $a$ and 
$1\leq n\leq M$ for $b$) plus a common normalisation factor $h_{DD}$ in the system of 
equations (\ref{eq:abcd_split}). Nevertheless, the statistical precision in this approach 
appears to be better as will be seen in the feasibility study. 

In principle, one can even consider splitting the Dalitz plot into more regions, 
but certainly the increase in the number of free parameters can diminish the 
possible gain in statistical precision. 
Any strategy involving splitting the Dalitz plot should be optimised taking into account the 
size of experimentally available samples of correlated $\Dz\Dzb$ and \bdk decays. 

\section{Simulation results}

\label{sec:simulation}

To test the feasibility of the proposed method, simulation studies using pseudoexperiments are performed. 
Samples of flavour-specific \dkpp decays, correlated $\Dz\Dzb$ pairs decaying to \kpp, and 
\dkpp decays from \bdk are generated using the $D$ decay amplitude measured by Belle 
collaboration~\cite{Poluektov:2010wz}. Samples are simulated with 
$r_B=0.1$, $\gamma=60^{\circ}$ and $\delta_B=130^{\circ}$ which is close to the results of the 
recent model-independent measurement of the \bdk, \dkpp channel by the LHCb collaboration~\cite{Aaij:2014uva}.
For each of those event samples, the Fourier series coefficients and their covariance matrices 
are calculated 
as described in Section~\ref{sec:fourier-non-split}. Systems of equations which 
contain relations between Fourier spectrum coefficients of flavour-specific 
$D$, $\Dz\Dzb$ and \bdk densities are then solved by maximising the combined likelihood
to obtain the value of $\gamma$. 

The formalism in Sections~\ref{sec:fourier-non-split} and \ref{sec:fourier-split} involved Cartesian 
\CP-violating parameters $x_{\pm}$ and $y_{\pm}$. This approach is likely more suitable when dealing 
with real data when one has to combine the results of different $\gamma$-sensitive analyses. In the 
simulation study presented here, the free parameters are chosen to be $\gamma$, $r_B$ and $\delta_B$. 

For the flavour-specific \dkpp mode, a large sample of $10^7$ generated events is used. 
This sample is not expected to contribute significantly to the uncertainty on $\gamma$ 
since high-statistics data sets are available at both the $B$ factories and LHCb. 
The size of the \bdk, \dkpp sample generated is $10^4$ events for each $B$ meson flavour, 
which corresponds roughly to $10$ times the data sample from LHCb Run 1~\cite{Aaij:2014uva}. 
Three scenarios with different correlated $\Dz\Dzb$ sample sizes are considered, 
$10^5$, $10^4$ and $10^3$ events. For comparison, the $e^+e^-\to \Dz\Dzb$ data sample collected by 
CLEO experiment where both $D$ mesons decay into \kpp contains 473 events, 
however, many other $D$ decay modes are used in the combined fit to obtain the phase coefficients
(notably, the modes where one of the $D$ mesons is reconstructed in a \CP eigenstate or 
as $\KL\pip\pim$)~\cite{Libby:2010nu}. It is expected that the statistical uncertainty of the 
$\Dz\Dzb$ sample of $10^5$ events will contribute negligibly to the uncertainty on 
$\gamma$, thus pseudoexperiments with this sample size probe uniquely how the 
approximation of the amplitude with a finite number of parameters 
(\ie truncated Fourier series or limited number of bins) affects $\gamma$
sensitivity. The low-statistics sample of $10^3$ events, on the other hand, will demonstrate the 
contribution of a limited $\Dz\Dzb$ sample to the sensitivity. 

Each ensemble of pseudoexperiments is 
fitted with the binned model-independent procedure with $3,5,8,12$, and $20$ bins 
using both the phase-difference and ``optimal'' binning schemes~\cite{Bondar:2008hh}, 
and with the two Fourier analysis techniques outlined above, using the entire Dalitz plot 
or the Dalitz plot split in two regions, respectively. In the approaches with Fourier expansion, 
the limit $M$ on the number of harmonics is set to $M=1, 2, 4, 7, 11$, or $19$. 
In addition, an unbinned model-dependent fit is performed to serve as 
a reference for the best possible statistical $\gamma$ precision that can be reached. 

\begin{figure}
\centering
\includegraphics[width=0.4\textwidth]{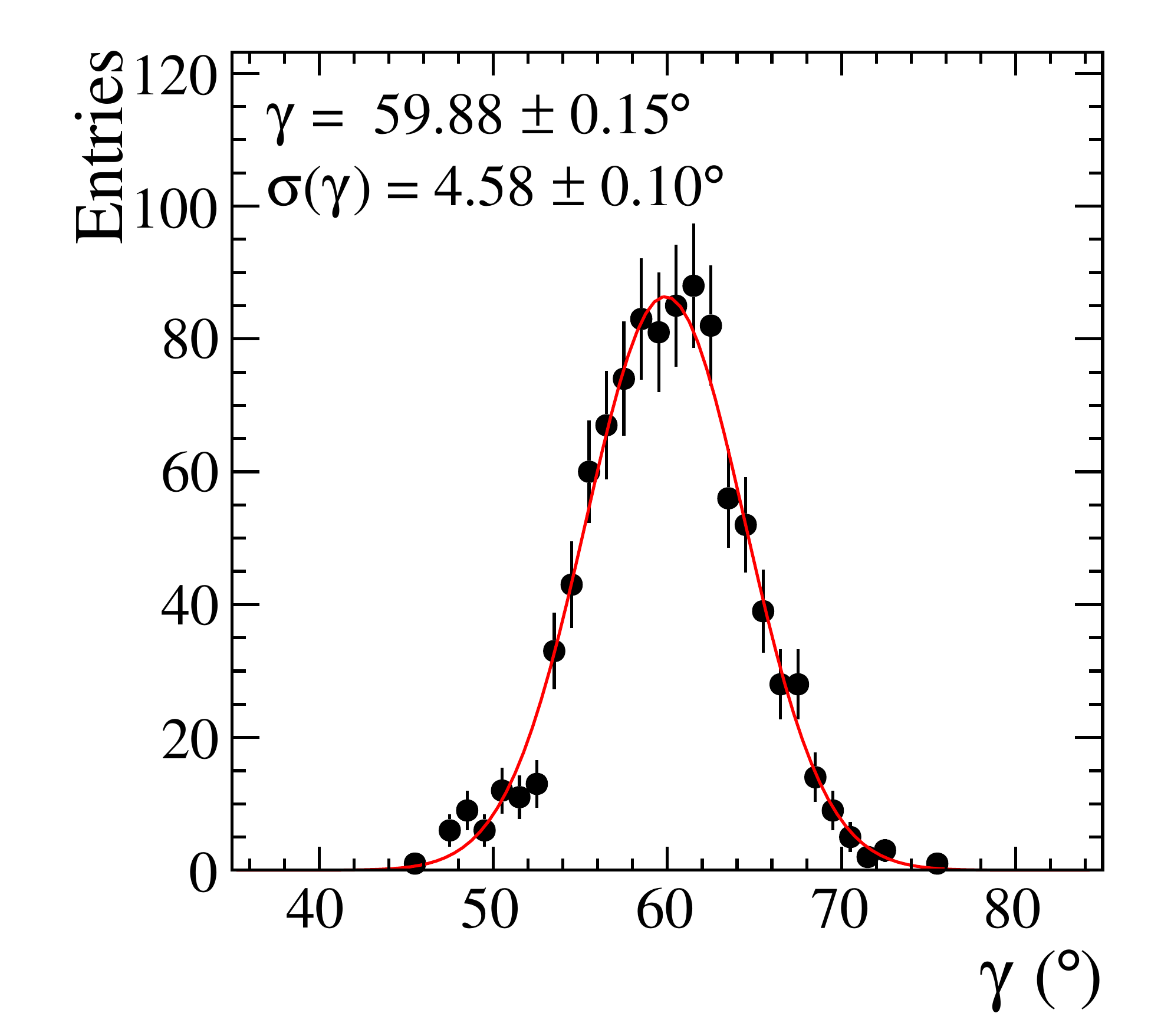}
\put(-35, 135){(a)}
\includegraphics[width=0.4\textwidth]{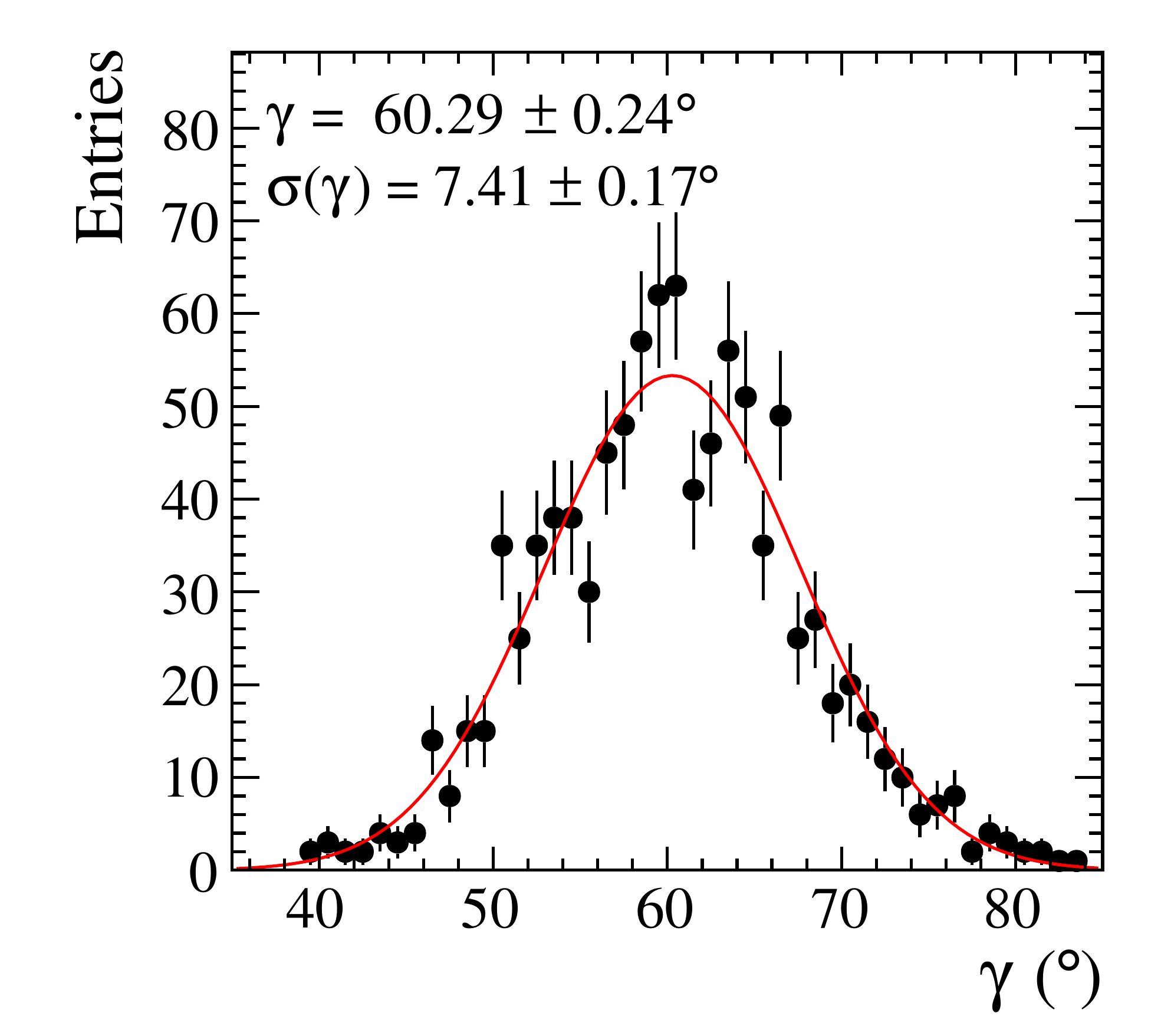}
\put(-35, 135){(b)}
\caption{Distributions of the reconstructed value of $\gamma$ in pseudoexperiments 
         with the (a) baseline and (b) reduced \dkpp model used for phase 
         difference calculation. Points are histograms of fit results, 
         and the solid red line is the result of a fit with Gaussian 
         distribution. The numerical results of the Gaussian fit are 
         also reported. }
\label{fig:resid}
\end{figure}

The Fourier expansion approach is verified to produce unbiased results if different 
\dkpp amplitudes are used for event generation and calculation of the phase difference 
$\Phi(\dvar)$. This is certainly a requirement for a technique to be model-independent. 
This check is performed by using a reduced \dkpp model where a subset of two-body 
amplitudes is present ($\rho(770)^0$, $\omega(782)$ and $f_0(980)$ in the 
$\pip\pim$ amplitude, and $K^*(892)^{\pm}$ in the $\KS\pi$ amplitudes) 
plus a flat non-resonant term. The results in Fig.~\ref{fig:resid} are shown for $M=1$ and 
$10^5$ $\Dz\Dzb$ events, but a similar check is performed for each value of $M$. 

\begin{figure}
\centering
\includegraphics[width=0.4\textwidth]{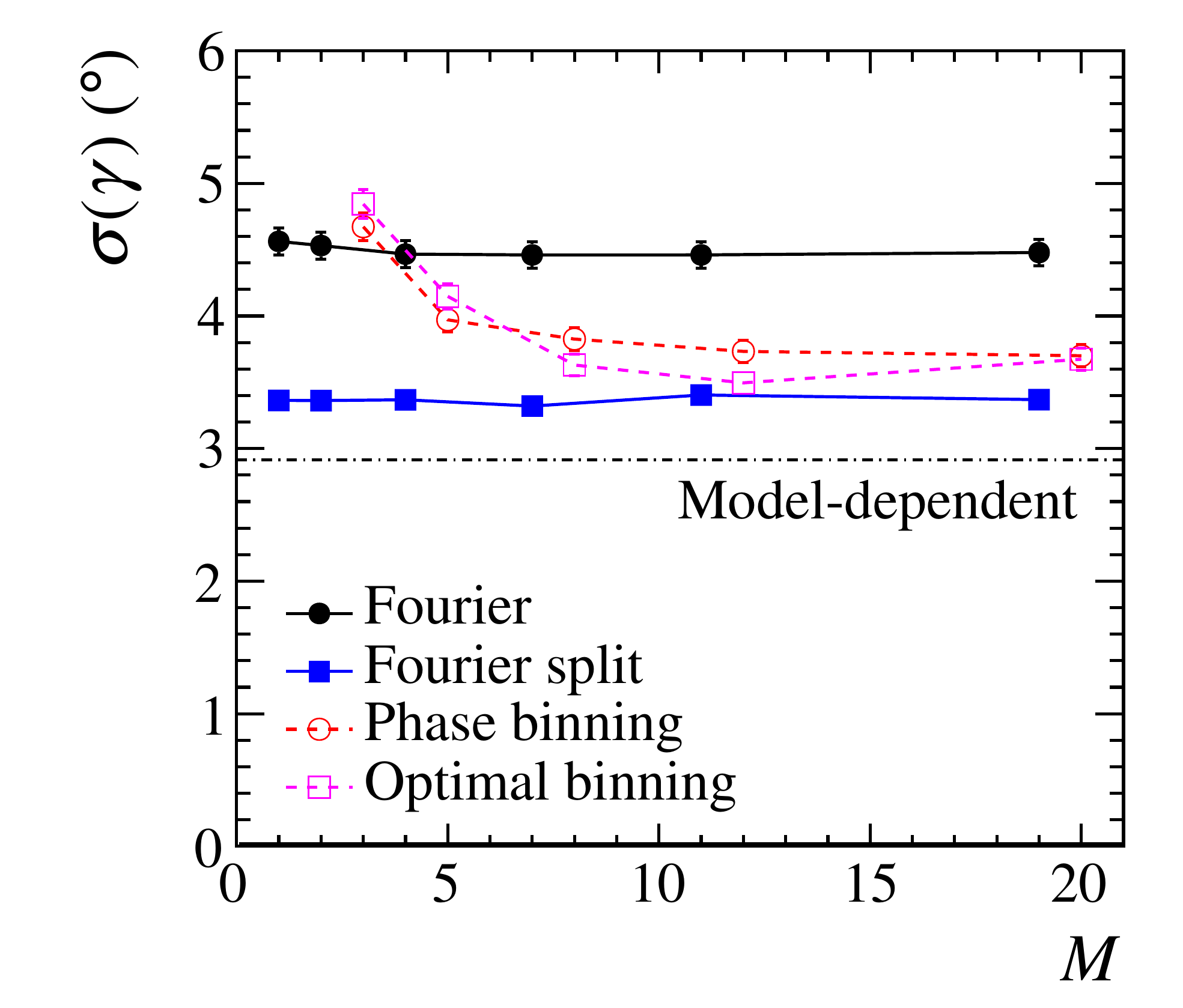}
\put(-30, 35){(a)}
\includegraphics[width=0.4\textwidth]{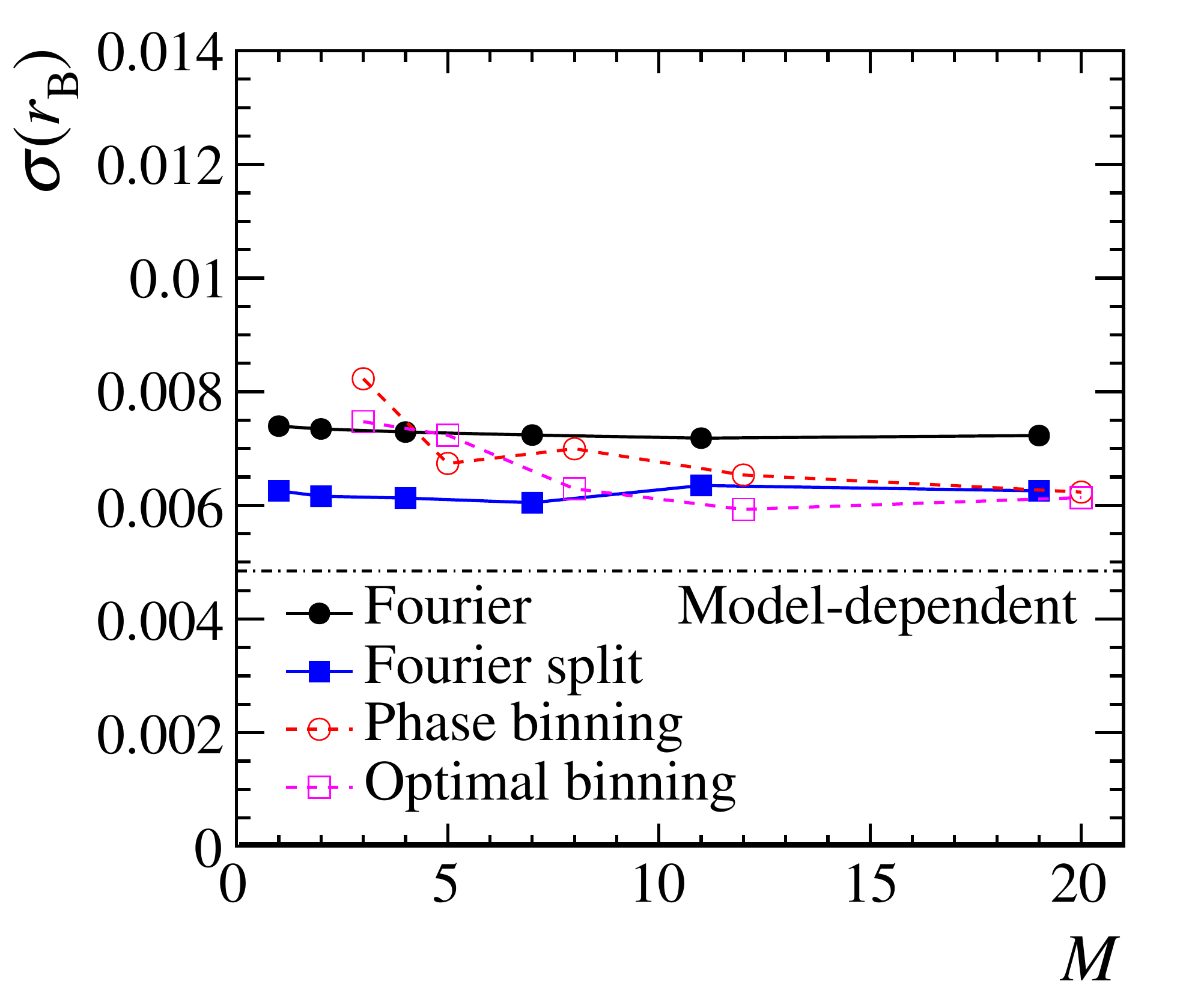}
\put(-30, 35){(b)}

\includegraphics[width=0.4\textwidth]{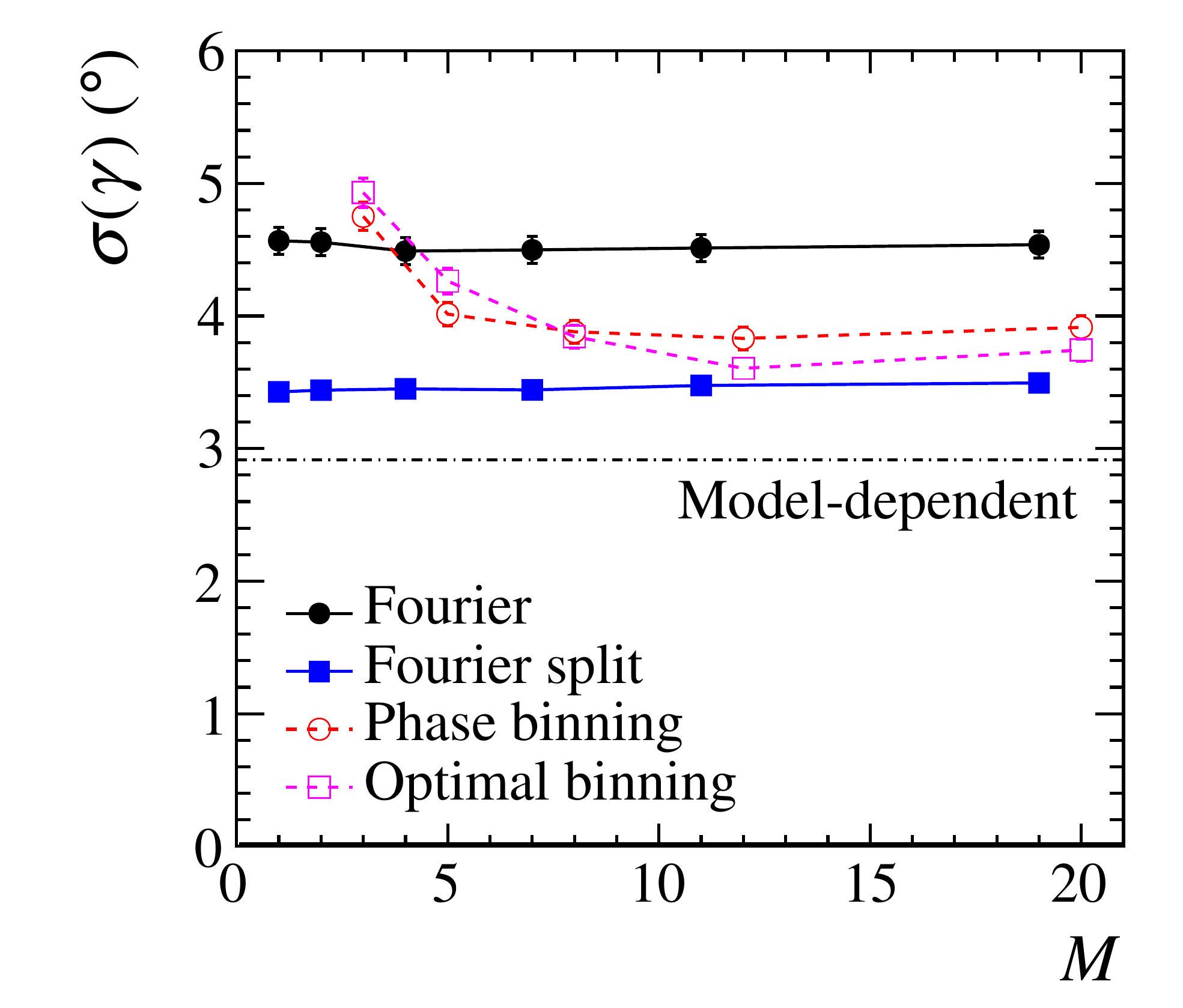}
\put(-30, 35){(c)}
\includegraphics[width=0.4\textwidth]{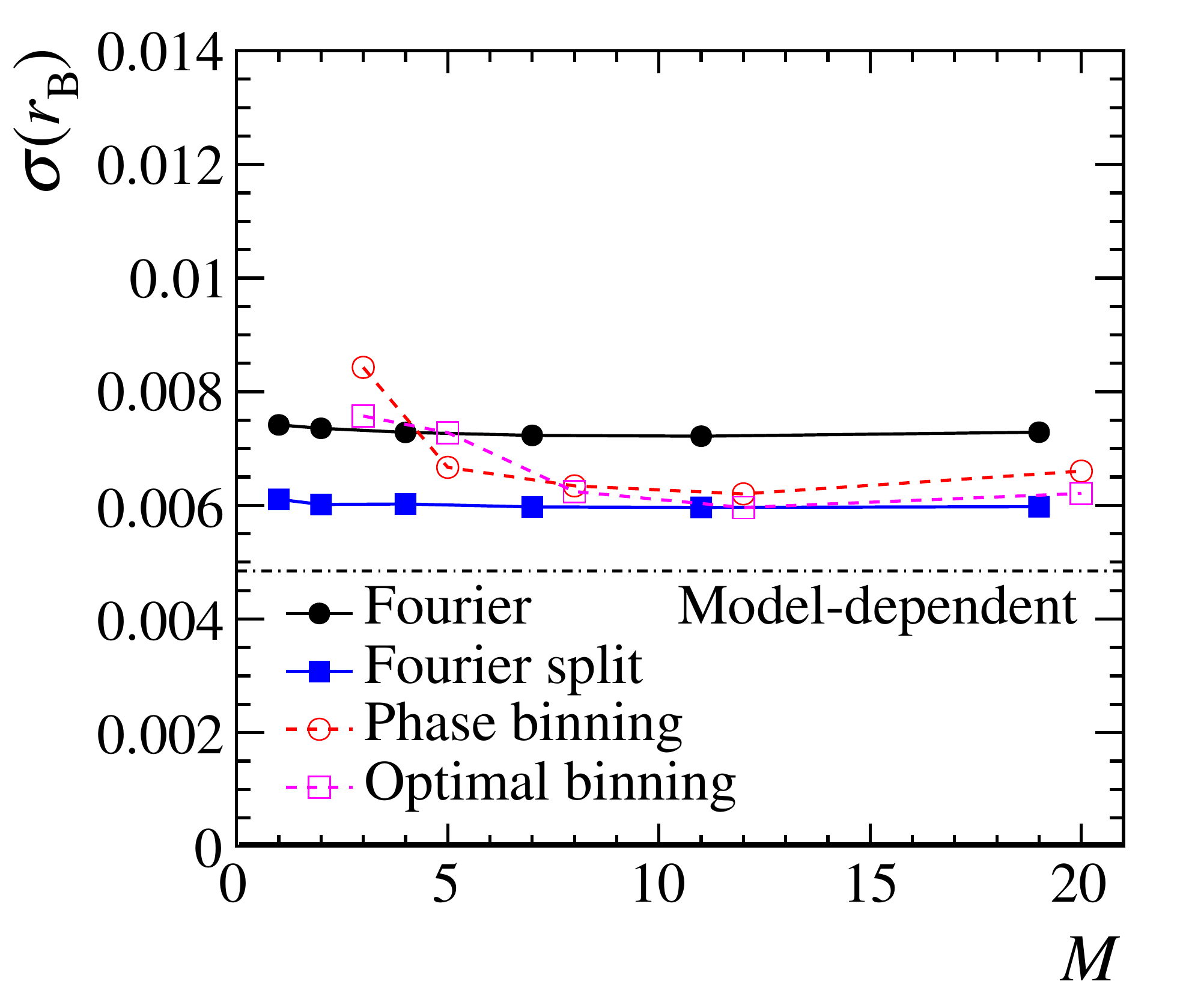}
\put(-30, 35){(d)}

\includegraphics[width=0.4\textwidth]{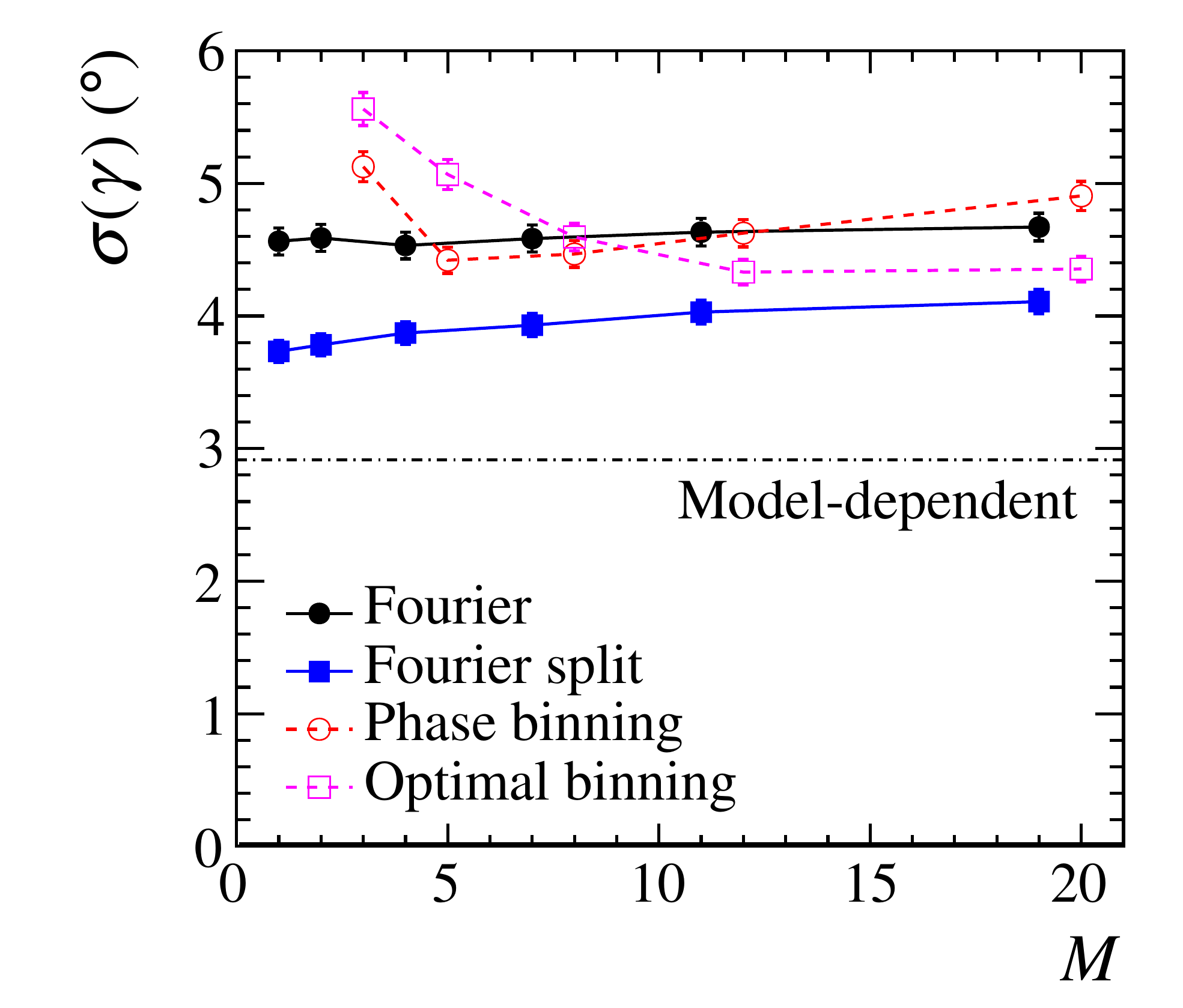}
\put(-30, 35){(e)}
\includegraphics[width=0.4\textwidth]{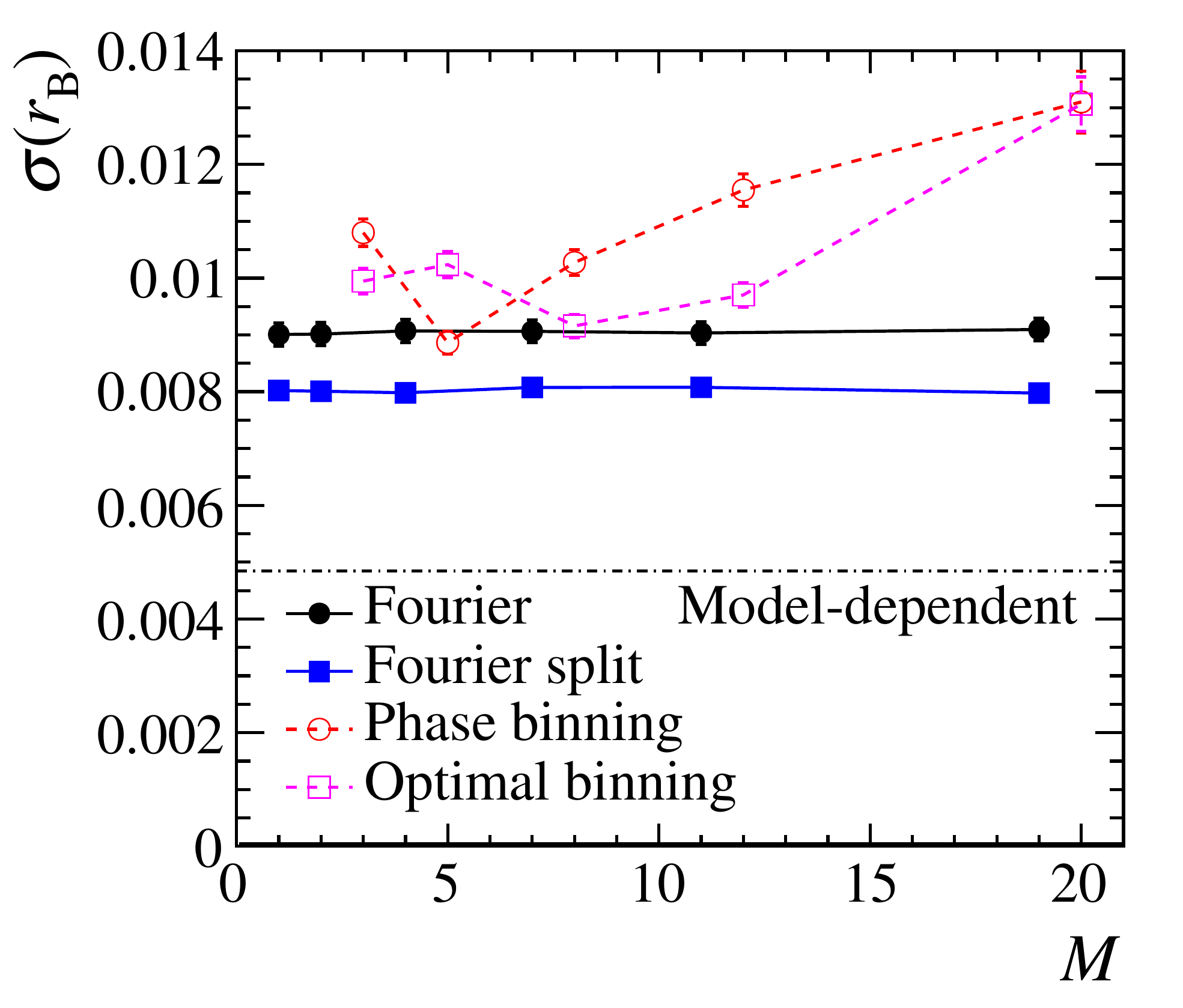}
\put(-30, 35){(f)}

\caption{Statistical precision of (a,c,e) $\gamma$ and (b,d,f) $r_B$ measurement 
         as a function of the number of Fourier terms (in the unbinned) 
         or number of bins (in the binned approach) for different model-independent 
         fit strategies, and its comparison with the model-dependent unbinned fit. 
         The generated $\Dz\Dzb$ sample size is (a,b) $10^5$, (c,d) $10^4$ and (e,f) $10^3$ events. }
\label{fig:res}
\end{figure}

Figure~\ref{fig:res} shows the $\gamma$ and $r_B$ resolutions as functions of the number of bins 
(for the binned scenarios) and the number of Fourier expansion terms 
(for the unbinned scenarios) with the four fit strategies described above and for the three different 
$\Dz\Dzb$ sample sizes. For comparison, the uncertainty of the unbinned model-dependent fit 
is also shown. While the precision of the binned approaches depends on the number of bins, 
the uncertainty of the Fourier expansion techniques practically does not depend on the number of 
harmonics $M$ for relatively large $\Dz\Dzb$ samples sizes, while for a small 
$\Dz\Dzb$ sample size of $10^3$ the optimum is reached for $M=1$ ({\it i.e.} for the smallest 
possible number of free parameters, which is three for non-split and six for split Dalitz plot). 
It is possible that other multibody $D$ decays may require higher harmonics to reach optimal 
sensitivity. Another case when Fourier terms with $n>1$ might be required is if the amplitude 
model $A_D^{\rm (model)}(\dvar)$ used to define $\Phi(\dvar)$ differs significantly from the true 
one. 

\begin{table}
  \caption{Uncertainty of $\gamma$ measurement with strategies using binned fit (with optimal binning)
           and using Fourier expansion (with non-split and split Dalitz plot). 
           The numbers correspond to the best 
           $\gamma$ resolution obtained in a range of $M$ (see Fig.~\ref{fig:res}). 
           For comparison, the $\gamma$ uncertainty
           for unbinned model-dependent fit is $\sigma(\gamma)=2.91\pm 0.07^{\circ}$. }
  \label{tab:res}
  \centering
  \begin{tabular}{l|ccc}
    \hline
    Sample size & \multicolumn{3}{c}{$\gamma$ resolution, ${}^{\circ}$} \\
                  \cline{2-4}
                & Binned optimal & Fourier non-split & Fourier split \\
    \hline
    $10^4$ \bdk, $10^3$ \ddb & $4.33\pm 0.10$ & $4.54\pm 0.10$ & $3.73\pm 0.08$ \\
    $10^4$ \bdk, $10^4$ \ddb & $3.60\pm 0.08$ & $4.51\pm 0.10$ & $3.43\pm 0.08$ \\
    $10^4$ \bdk, $10^5$ \ddb & $3.49\pm 0.10$ & $4.47\pm 0.10$ & $3.32\pm 0.08$ \\
    \hline
  \end{tabular}
\end{table}

The $\gamma$ uncertainties for the optimal scenarios with the binned and unbinned techniques 
are compared in Table~\ref{tab:res}. The uncertainty of the approach with split Dalitz plot is 
significantly better than when the Dalitz plot is taken as a whole. 
It is also clear that the Fourier expansion technique with split Dalitz plot shows 
better sensitivity than the binned method using ``optimal'' binning, with the gain being the 
most significant for smaller $\Dz\Dzb$ sample size. The technique, however, 
is still about 10\% less sensitive than the unbinned model-dependent approach. The 
possibilities to further improve the sensitivity of the unbinned model-independent 
method are discussed in Section~\ref{sec:improvements}.

\section{Practical considerations}

\label{sec:practical}

To be applicable to real data, the technique should be able to deal with experimental effects 
such as backgrounds and non-uniform detection efficiency across the Dalitz plot. 
Since background enters the decay density additively, it can be treated at the level of 
Fourier-transformed variables, by calculating the Fourier expansion of the background
density and subtracting it from the coefficients calculated on data. 
On the other hand, efficiency enters the density in a multiplicative way, 
thus Fourier expansion need to be applied to efficiency-corrected data. 
The correction can be applied on an event-by-event basis, 
by assigning each event a weight proportional to the inverse of efficiency 
while calculating the Fourier coefficients. 

The studies presented above have been performed using a combined likelihood fit to both 
$\bdk$ and correlated $\Dz\Dzb$ samples. It is also possible
to perform the analysis in two stages, by first calculating the coefficients of Fourier 
transformation of the functions $C(\phi)$ and $S(\phi)$ from the $\Dz\Dzb$ data, followed 
by a fit to \bdk sample using the coefficients, their correlations and uncertainties from the 
first stage. This is likely to be more convenient in practice, since the data samples come 
from different experiments.

\section{Further directions of development}

\label{sec:improvements}

Using notation of the generalised model-independent formalism presented in Section~\ref{sec:general}, 
the Fourier analysis technique proposed above uses a family of $2M+1$ weight functions
\begin{equation}
  \begin{split}
    w_{0}(\dvar) & = 1, \\
    w_{n}(\dvar) & = \cos[n\Phi(\dvar)], \\
    w_{n+M}(\dvar) & = \sin[n\Phi(\dvar)], \\
  \end{split}
\end{equation}
where $1\leq n\leq M$. The use of the function $\Phi(\dvar)$ ensures that different points in the phase space 
do not cancel each other out while calculating the integral, and thus the interference term that provides 
sensitivity to \CP-violating observables is large (assuming, of course, that the amplitude model that 
provides $\Phi(\dvar)$ is close to the true amplitude). However, information about the absolute 
value of the amplitude is ignored in the formalism presented in Section~\ref{sec:fourier-non-split}
and is taken into account only rather roughly in Section~\ref{sec:fourier-split}. Alternatively, 
one could consider a weight function that in addition takes into account the magnitudes of the favoured and suppressed amplitudes from the model $|\overline{A}^{\rm (model)}_D(\dvar)|$ and 
$|A^{\rm (model)}_D(\dvar)|$, and thus adds more information to maximise the interference term. 
In the presence of background, the family of weight functions should also take into account the 
distribution of background events over the phase space. Further optimisation of the 
family of weight functions needs additional study. 

The proposed technique could be especially useful in cases where a binned approach will limit precision due to small sample sizes of decays which determine the phase information. 
Examples are the $\dkkk$ mode, where the sample of quantum-correlated decays is small and currently only two bins are used in the $\gamma$ measurement~\cite{Aaij:2014uva}. 
Another example is \bdkpi decays, where the phase coefficients corresponding 
to the three-body $B$ decay are free parameters together with $\gamma$~\cite{Gershon:2009qr, Craik}. 
Having an amplitude model which describes the strong phase variation across the $B$ decay Dalitz plot 
with a small number of parameters should improve the statistical sensitivity. 

Other analyses, where the coherent $\Dz-\Dzb$ admixtures are involved, 
are measurements of charm mixing and \CP violation in mixing, as well as measurement of the UT angle $\beta$ in 
$B\to D h^0$ decays. These classes of measurements 
utilise oscillations of $\Dz$ and $\Bz$ mesons, respectively, and thus the parameters of the 
$\Dz-\Dzb$ admixture are functions of decay time. In the proposed formalism, the coefficients 
of the Fourier series will be functions of decay time as well. While such analyses will 
certainly be more complicated than the case with constant coefficients, they are 
conceptually similar to the measurements using the binned technique which have already been 
carried out~\cite{Aaij:2015xoa,Vorobyev:2016npn}. 

\section{Conclusion}

A technique to perform unbinned model-independent analysis of a coherent admixture of \Dz and \Dzb states decaying to a multibody final state is proposed. It is illustrated in detail using the measurement of the UT angle 
$\gamma$ from \bdk decays. Unlike the well-known technique with 
Dalitz plot binning, the proposed method employs Fourier analysis of the spectrum of the strong phase difference 
between the \Dz and \Dzb amplitudes. While the method relies on an amplitude model to reach optimal statistical 
precision, it is unbiased by construction even if the wrong model is used. 

A study of the feasibility of the proposed method has been performed with simulated pseudoexperiments. 
The precision of the method does not depend strongly on the number of Fourier expansion terms used, 
and even with only the single leading term yields sensitivity comparable to that of the 
binned model-independent approach. 
A modification of the procedure, where Fourier expansion is performed in two regions of the Dalitz plot 
separated according to the ratio of the suppressed and favoured amplitudes, provides $\gamma$ sensitivity 
better than the most optimal binned strategy. The gain compared to the binned approach is especially 
significant if the size of the correlated $\Dz\Dzb$ sample, which determines the strong phase in $D$ 
meson decay, is small. Possible ways of improving the sensitivity of the proposed technique even further 
are identified and need further study. 

The method is not limited to $\gamma$ measurements with three-body $D$ decays and can be generalised 
to any analysis where the parameters of a coherent admixture of $\Dz$ and $\Dzb$ in a multibody final state 
need to be determined, such as measurements of charm mixing and \CP violation, and measurements of the UT 
angle $\beta$ in $B\to D h^0$ decays. The technique could also be useful in $\gamma$ measurements with 
a double Dalitz plot analysis of \bdkpi, \dkpp decay; in that case the Fourier expansion can be applied 
to both the $B$ and $D$ Dalitz plots. 

\section*{Acknowledgements}

The author is grateful to his colleagues from the LHCb collaboration for stimulating discussions and 
help in preparing the paper: Timothy Gershon, Mark Whitehead, Guy Wilkinson, Wenbin Qian, Susan Haines, 
Matthew Kenzie, and other members of ``beauty decays to open charm'' analysis working group. 
The author would like to thank Alex Bondar for inspiring the search for model-independent 
approaches beyond well-developed binned technique. 
The work is supported by the Science and Technology Facilities Council (United Kingdom). 

\bibliographystyle{LHCb}
\bibliography{umig}

\end{document}